\documentclass[pre,twocolumn,showpacs,epsfig]{revtex4-1}
\usepackage{ulem}
\usepackage{graphicx}
\usepackage{graphics}
\usepackage{wrapfig}
\usepackage{dcolumn}
\usepackage{amssymb}
\usepackage{bm}
\usepackage{epsfig}
\usepackage{float}

\usepackage{psfrag}
\usepackage{color}
\usepackage[utf8x]{inputenc}
\usepackage{ucs}
\usepackage{amsmath}

\usepackage{amsfonts}
\usepackage{amssymb}
\usepackage{graphicx}
\usepackage{dcolumn}
\usepackage{bm}
\usepackage{latexsym}
\usepackage{hyperref}

\usepackage{xcolor}
\hypersetup{
    colorlinks,
    linkcolor={purple!50!black},
    citecolor={purple!50!black},
    urlcolor={purple!80!black}
}

\def\3dots{\:\raisebox{-0.5ex}{$\stackrel{\textstyle.}{:}$}\:}
\def\beq{\begin{equation}}
\def\eeq{\end{equation}}
\def\bea{\begin{eqnarray}}
\def\eea{\end{eqnarray}}

\begin{document}
\title{{Active nonreciprocal attraction between motile particles in an elastic medium.}}
\author{Rahul Kumar Gupta$^{1,2,3}$, Raushan Kant$^{2}$, Harsh Soni$^{2}$, A.K. Sood$^2$  and Sriram Ramaswamy$^{2}$}

\affiliation{$^1$Tata Institute of Fundamental Research, Gopanpally, Hyderabad 500 107 India}
\affiliation{$^2$Department of Physics, Indian Institute of Science, Bangalore 560 012, India}
\affiliation{$^3$ Institut für Theoretische Physik II - Soft Matter
	Heinrich-Heine-Universität, 40225 Düsseldorf, Germany}
\date{\today}
\begin{abstract} We show from experiments and simulations on vibration-activated granular matter that self-propelled polar rods in an elastic medium on a substrate turn and move towards each other. We account for this effective attraction through a coarse-grained theory of a motile particle as a moving point-force density that creates elastic strains in the medium that reorient other particles. Our measurements confirm {qualitatively the predicted features of the distortions created by the rods, including the $|x|^{-1/2}$ tail of the trailing displacement field and non-reciprocal sensing and pursuit. A discrepancy between the magnitudes of displacements along and transverse to the direction of motion remains.} Our theory should be of relevance to the interaction of motile cells in the extracellular matrix or in a supported layer of gel or tissue.
\end{abstract}
\maketitle

\section{Introduction}
Active systems with local conservation laws generically display long-range interactions. Swimmers in a viscous fluid interact hydrodynamically, with profound consequences for pair \cite{YeomansPrl2007,baskaran2009statistical} and collective behaviour \cite{simha2002hydrodynamic,SriramRMP,saintillan2015theory,elgeti2015physics}. Autophoretic colloids communicate through their diffusive chemical fields \cite{golestanian2019phoretic}, and passive inclusions through an ambient conserved active-particle density \cite{granek2020bodies}. A broken-symmetry mode leads to richer interactions, for example through 
suppression of nonequilibrium \cite{bartolo2003effective} fluctuations, or strain fields generated by non-motile force dipoles  
\cite{SafranPrl2002,SafranPre2004,SafranPre2013,safranSoftM2016,golkov2017shape,golkov2019elastic}. Active \textit{polar} order in elastic media has been studied in
\cite{marcq2014spatio,banerjee2011substrate,banerjee2015propagating,kopf2013non,notbohm2016cellular,notbohm2016cellular,alaimo2016microscopic,menzel2013traveling},
with a fully rotation-invariant theory in \cite{maitra2019oriented}. Henkes et al. \cite{henkes2020dense} study the effect of a collection of motile but \textit{noninteracting} active Brownian particles on an elastic medium on a substrate. 

In this paper we study the effects of self-propulsion in a crystalline medium with emphasis on single-particle dynamics and pair interactions. We present findings from experiments and simulations on rods tapered towards one tip which acquire motility by transducing the energy of a vibrating substrate and move through a dense monolayer of beads. We account for our observations through a theory whose further predictions we test.  

\begin{figure}
	\begin{center}
		\includegraphics[width=0.48\textwidth]{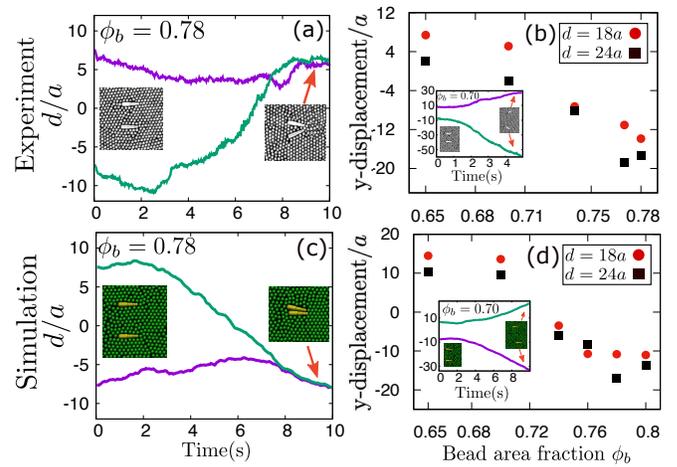}\\
			\caption{rods attract in an elastic medium: (a) \& (c) show, in experiment and simulation respectively, the $y$ coordinates of a pair of rods initially positioned as in the insets, and separated only along $y$ by $d_0=15a$, where $a$ is the bead radius, at bead area fraction $\phi_b=0.78$ deep in the crystalline phase of the bead layer. Experiment (b) \& simulation (d) show the change in the $y$ separation of the rods on traversing half the system size {along $x$, averaged} over 50 (12) independent runs in simulation (experiment) for $d_0=18a$ \& $d_0=24a$, as a function of $\phi_b$. The sign-change from positive to negative as $\phi_b$ increases past $0.75$ signals the transition from avoidance in the fluid phase [see insets to (b) and (d)] to effective attraction upon entering the crystalline phase.} \label{trajectory}
	\end{center}
\end{figure}

Here are our main results. The polar rods retain their motility even in the crystalline phase that occurs at high bead coverage. In the crystal, a pair of parallel rods placed side-by-side turn and locomote \textit{towards} each other [Fig. \ref{trajectory} {(a) and (c)}] -- unlike in the fluid phase at lower coverage [Fig. \ref{trajectory} (b) and (d), inset]. We account for this behavior through a theory of a damped elastic medium driven by moving point forces carried by the motile rods, whose orientations respond to the strains of the medium (Fig. \ref{attraction}). Our theory predicts, and our numerical measurements confirm, a distinctive form of the elastic distortion created by a single moving rod, decaying exponentially ahead of the rod and as (distance)$^{-1/2}$ directly behind it (Fig. \ref{ufield}).
Our theory further implies a non-reciprocal  
\cite{das2002driven,das2004nonequilibrium,ivlev2015statistical,cavagna2017nonsymmetric,chen2017fore,saha2019pairing,dadhichi2020nonmutual,fruchart2020phase,saha2020scalar,you2020nonreciprocity} 
pursuit-and-capture interaction between the rods, which our experiments and simulations reproduce (Fig. \ref{nonreciptheory}).

The rest of this paper is organized as follows. In section II we describe experimental and simulation details of our system. Section III demonstrate characterization of bead medium in terms of hexatic order parameter and microrehology analysis. In section IV we present results on bead velocity field and two particle attraction.  Section V presents theory of motile particle in an isotropic elastic medium and compare the results extracted from the simulation. Section VI presents our results on non-reciprocal interaction and discusses the effect of polar coupling terms on the particle dynamics, and we end with a brief Conclusion in VII.

\section{Experimental and Simulation details}
We work with a well-established model active-matter system \cite{NKPRL, NKPRE, NatCom, trapping, harshsoftmatter}: brass rods, $4.5$ mm long and tapered towards one end and hence geometrically polar,  with diameters 1.1 mm and 0.8 mm at the two ends, surrounded by aluminium spheres of radius $a=0.4$ {mm, the whole} confined between a circular base plate of diameter 13 cm and a glass lid, separated by a gap $w=$1.2 mm. The plate is mounted on a permanent-magnet shaker (LDS V406-PA 100E) driving it sinusoidally in the vertical direction with frequency $f$= 200 Hz and shaking strength $\Gamma \equiv {\mathcal{A}(2 \pi f)^{2}}/{g} = 7.0$, where $\mathcal{A}$ and $g$ are the respective shaking amplitude and the gravitational acceleration. The vertical vibration endows the rods with motility through mechanical asymmetry and static friction \cite{narayan2010phase,yamada2003coherent}. Images of the particles are recorded using a high-speed camera (Redlake MotionPro X3), which are further processed using ImageJ~\cite{ImageJ} to extract the position, velocity and orientation of the particles.

In numerical simulations, the rod is modelled as an array of overlapping spheres of different sizes ~\cite{NatCom} and the vertically vibrating plate and glass lid are represented by the two horizontal walls whose vertical positions at time $t$ are given by $\mathcal{A} \cos 2 \pi f t$ and $\mathcal{A} \cos 2\pi f t + w$, respectively. The particle-particle and particle-wall collisions are governed by the Impulse-based collision model~\cite{stronge} and the ballistic dynamics of the particles is governed by Newtonian rigid body dynamics. We use the time-driven particle dynamics algorithm to perform the simulations.  VMD software~\cite{VMD} is used to make all the movies and snapshots from the simulations. The values of the friction and restitution coefficients are 0.05 and 0.3 for particle-particle collisions, 0.03 and 0.1 for rod-wall collisions, and 0.2 and 0.3 for bead-wall collisions, respectively. Simulations are done in a square box of side length 78$a$, with periodic boundary conditions applied in the $xy$ plane.

\section{Characterization of the bead medium}
\subsection{Hexatic order parameter and its correlation function}
\begin{figure}[htb]
	\begin{center}
		\includegraphics[width=0.5\textwidth]{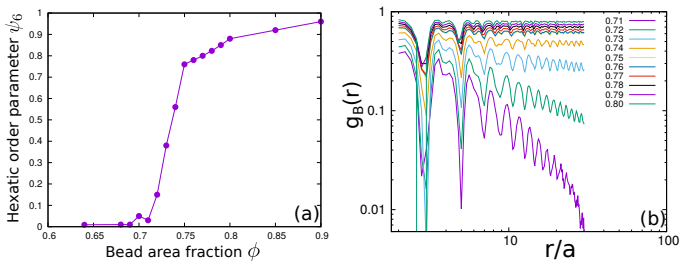}\\
		\caption{(a) $\psi_6$ order parameter as a function of bead area fraction $\phi_b$. (b) Bond orientational correlation function vs distance $r$ scaled by bead radius $a$ at different values of $\phi_b$.}
		\label{hop}
	\end{center}
\end{figure}

We examine here the ordering of the bead medium without rods \cite{olafsen2005two}, as a function of bead area fraction $\phi_b$. We focus on the global 6-fold bond-orientational order parameter $\psi_6$ and the bond-orientational correlation function $g_B(r)$. {In view of the close correspondence between the behavior of vibrated-sphere monolayers \cite{olafsen2005two} and equilibrium 2-stage melting \cite{KTHNY2} we use the properties of $\psi_6$ as diagnostics of crystalline as well as hexatic order.} We first evaluate the local orientational order parameter  
\begin{equation}
\psi_{6,i} = \frac{1}{N_i} \sum_{j=1}^{N_i} \exp(6i\theta_{ij}),
\end{equation} 
where $N_i$ is the number of particles within a cut-off distance $r_{\text{min}}$ from the $i^\text{th}$ particle. $\theta_{ij}$ is the angle made by the vector from the centre of the $i^\text{th}$ to that of the $j^\text{th}$ bead with respect to a reference direction, and $r_{\text{min}}$ is the position of the first minimum of the radial distribution function
\begin{equation}
g(r)=\dfrac{1}{\rho_0}\langle\rho(\mathbf{r}') \rho(\mathbf{r}'')\rangle_{r}.
\end{equation}
In the above expression, the angular bracket stands for the ensemble average over all values of $\mathbf{r}'$ and $\mathbf{r}''$  with $\left| \mathbf{r}'-\mathbf{r}''\right| =r$, 
\begin{equation}
\rho(\mathbf{r})=\sum^N_{i=1} \delta(\mathbf{r}-\mathbf{r}_i)
\end{equation}
is the number density field, and $\rho_0$ is the average number density.
We then calculate  the global bond orientational order parameter
\begin{equation}
\psi_6 = {1 \over N}\Biggl\lvert \sum_{i=1}^{N} \psi_{6,i} \Biggr\rvert.
\end{equation}
{Fig.~\ref{hop}a shows $\psi_6$ as a function of $\phi_b$, with a clear onset of nonzero $\psi_6$ around $\phi_b=0.72$, providing evidence that the bead medium is an isotropic liquid for $\phi_b<0.72$.} Above $\phi_b=0.72$, $\psi_6$ grows gradually with area fraction $\phi_b$ and saturates for $\phi_b>78$. 

{Measurements of the global $\psi_6$ for a single system size are insufficient to tell us about phases and transitions. However, clear indications of behaviour consistent with 
	two-stage melting \cite{KTHNY2} are seen in 
	the bond orientational correlation function}
\begin{equation*}
g_B(r) = \frac{\bigg< \psi_{6}^*(\mathbf{r'})\psi_{6}(\mathbf{r}'') \bigg>_{\left| \mathbf{r}'-\mathbf{r}''\right| =r}}{g(r)}. 
\end{equation*}
Fig.~\ref{hop}b illustrate that $g_B(r)$ decays exponentially for $\phi_b<0.72$, and for $0.72 \le \phi_b \le 0.75$ it decays algebraically with $r$. For $\phi_b>0.75$, $g_B(r)$ approaches a nonzero value for large $r$. This is consistent with the 2D melting scenario \cite{KTHNY2} in which the crystalline phase has long-range bond order (and quasi-long-range translational order), and the hexatic phase has quasi-long-range bond order. We can therefore use a nonzero hexatic order parameter as a reliable identifier of the crystalline phase.

\subsection{Viscoelastic properties of the bead  medium}
We employ microrheology to explore the viscoelastic properties of the bead medium. We assume the force on a tagged particle of the bead medium at time $t$ is given by $-\int^t_0 k(t-t')(\mathbf{x}(t')-x(0))dt'$, {with a memory kernel $k(t)$ that we have chosen, for convenience, to act on the displacement rather than the velocity}. The equation of motion for the bead then reads 
\begin{equation} 
\label{beadeom}
m \dfrac{d\mathbf{v}}{dt}=-\Gamma \mathbf{v}-\int^t_0 k(t-t')\mathbf{x}(t')dt'+\bm{\eta}(t).
\end{equation}
Here $m$ is the mass of the bead, $-\Gamma \mathbf{v}(t)$ is the dissipative force acting on the bead due to the substrate and $\bm{\eta}(t)$ is noise force with zero mean due to the surrounding medium. For a purely elastic and a purely viscous medium, $k(t)\sim\delta(t)$ and $-d{\delta}(t)/dt$ respectively. Denoting Laplace transforms by a tilde, equation  \eqref{beadeom} becomes  
\begin{equation}\label{mr2}
m [s\tilde{\mathbf{v}}(s)-\mathbf{v}(0)]=-\Gamma \tilde{\mathbf{v}}-\tilde{k}(s)\left[ \tilde{\mathbf{x}}(s)-\dfrac{\mathbf{x}(0)}{s}\right] +\tilde{\bm{\eta}}(s), 
\end{equation}
which simplifies, using $\tilde{\mathbf{v}}(s)=s\tilde{\mathbf{x}}(s)-\mathbf{x}(0)$, to
\begin{equation}\label{mr3}
\tilde{\mathbf{v}}(s)=\dfrac{m\mathbf{v}(0)+\tilde{\bm{\eta}}(s)}{ms+\Gamma+\tilde{k}(s)/s}.
\end{equation}
The sample average of the dot product of the above equation with $\mathbf{v}(0)$ gives
\begin{equation}\label{mr3}
\left\langle \tilde{\mathbf{v}}(s)\cdot\mathbf{v}(0)\right\rangle  =\dfrac{m\left\langle\mathbf{v}^2(0)\right\rangle}{ms+\Gamma+\tilde{k}(s)/s}.
\end{equation}
Here we have used{, for $s>0$,} $\left\langle\mathbf{v}(0) \cdot \tilde{\bm{\eta}}(s)\right\rangle=\mathbf{v}(0)\cdot \left\langle\tilde{\bm{\eta}}(s)\right\rangle=0$ as $\left\langle\eta(t)\right\rangle=0$. Assuming an effective thermal description with temperature $T$ for the bead fluctuations in the absence of polar rods, implies 
\begin{equation}\label{mr3}
\left\langle \tilde{\mathbf{v}}(s)\cdot\mathbf{v}(0)\right\rangle  =\dfrac{2T}{ms+\Gamma+\tilde{k}(s)/s}
\end{equation}

As the Laplace transform of the mean square displacement $\Delta (t)=\left\langle\left| \mathbf{x}(t)-\mathbf{x}(0)\right| ^2\right\rangle$ is $\tilde{\Delta}  (s)=(2/s^2)\left\langle \tilde{\mathbf{v}}(s)\cdot\mathbf{v}(0)\right\rangle$, {Eq. \eqref{mr3}} implies 
\begin{equation}
\tilde{\Delta}  (s)=\dfrac{4T}{s(ms^2+\Gamma s+\tilde{k}(s))}.
\end{equation}

\begin{figure}[htb]
	\begin{center}
		\includegraphics[width=0.5\textwidth]{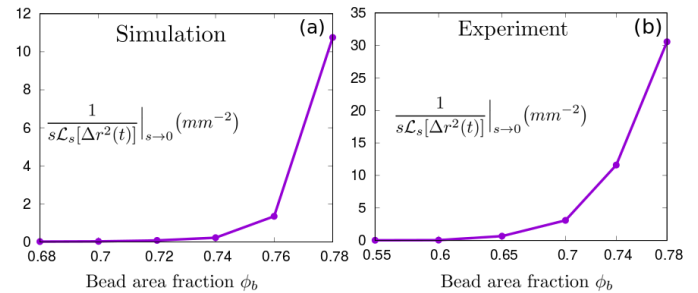}\\
		\caption{Stiffness parameter $1/(s \tilde{\Delta} (s))|_{s\to0}$ as a bead area fraction $\phi_b$ in simulation (a) and in Experiment (b).}
		\label{micro}
	\end{center}
\end{figure} 
We calculate the quantity $1/(s \tilde{\Delta}  (s))|_{s\to0}$ to estimate the stiffness of the effective potential felt by the bead. Fig.~\ref{micro} shows that $1/(s \tilde{\Delta}  (s))|_{s\to0}$ {is consistent with zero at small values of $\phi_b$, and shows an onset to nonzero values around $\phi_b = 0.7$. The onset $\phi_b$ is higher in the simulations than in the experiments, but the picture is broadly consistent with the presence of an elastic medium at large enough area fractions}.


\section{Results}
We now show key results from experiments and mechanically faithful simulations. In last section we establish, consistent with \cite{olafsen2005two}, the presence of liquid, hexatic and crystalline phases for bead area fraction $\phi_b<0.72$, $0.72 \le \phi_b < 0.75$ and $\phi_b \ge 0.75$ respectively and corroborate this structural characterization by single-particle microrheology on bead displacements, to establish the presence of a nonzero elastic stiffness at large enough $\phi_b$.

\begin{figure}
	\begin{center}
		\includegraphics[width=0.5\textwidth]{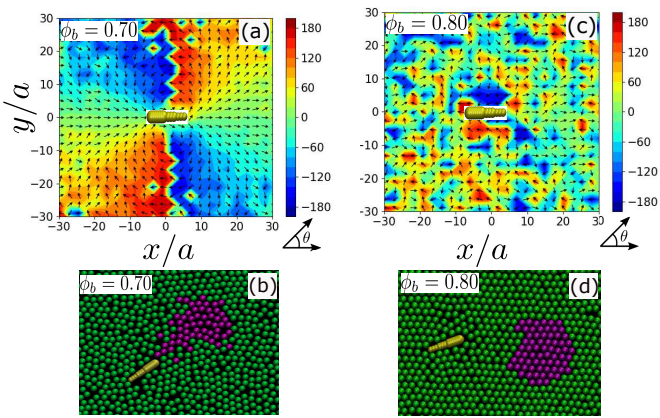}\\
		\caption{{
				Velocity field and dragging of beads: bead-velocity unit-vector field around a single motile rod at bead area fractions (a) $\phi_b = 0.70$ and (c) $0.80$. At $\phi_b = 0.70$, the flow pattern is similar to that of a monopole force density \citep{NatCom} in a fluid on a substrate; at $\phi_b = 0.80$, the velocity field is extinguished by the elasticity of the medium. Motile rod drags surrounding beads (marked purple) at (b) $\phi_b=0.70$, but leaves them undisturbed at (d) $\phi_b=0.80$ (see Supplementary Movie S5 and S6) \cite{supp}.}}
		\label{bead_drag}
	\end{center}
\end{figure}

In the liquid and hexatic phases the flow profile around a single rod ({Fig.~\ref{bead_drag}a \& \ref{bead_drag}b }) has the 
source-dipole form expected ~\cite{NatCom,brotto2013} for self-propulsion through a fluid on a substrate. 
In experiment and simulation in the liquid phase [e.g. at $\phi_b = 0.70$, Fig. \ref{trajectory}(b) \& (d) insets and Supplementary Movie S3, S4 \cite{supp}], a pair of polar rods initially pointing along, say, the $x$ direction, with initial separation $d_0$ strictly transverse to their orientation, turn away from each other and thus move apart, a simple consequence of the rotation of the orientation of each rod by the vortical flow generated by the other \cite{curlfoot}. When $\phi_b$ is increased past $0.75$ into the crystalline phase the polar rods remain motile but the large-scale flow they generate shuts down ({Fig.~\ref{bead_drag}c \& \ref{bead_drag}d }). 

Despite the absence of flow, a long-range pair interaction persists, but opposite in character to that in the fluid regime: the rods now turn to point and move \textit{toward} each other in both experiment and simulation (see Fig.~\ref{trajectory}a \& c for $\phi_b=0.78$ and $d_0 = 15a$ and Supplementary Movie S1 and S2 \cite{supp}). This effective attraction, more accurately a positive pairwise reorienting  or ``tactic'' \cite{o2020lamellar} response, depends crucially on the motility of the rods. 
Rods of the same central diameter, but tapered symmetrically at their two ends and hence non-motile \cite{VJScience}, show no detectable interaction when placed in the vibrated crystalline array of beads at area fraction $\phi_b=0.70$ and $\phi_b=0.80$. The apolar rods neither attract nor repel each other(see Supplementary Movie 7 and 8). This observation shows unambiguously that motility is crucial for the attractive interaction of polar rods in our studies. 
Static interactions between scalar or apolar inclusions such as those discussed in \cite{eshelby1957determination,katira}, if present, are insignificant compared to the interaction arising from motility-driven strain fields. 

As a measure of the nature of the interaction of the rods we use the relative transverse displacement $\Delta y$ of a pair of rods initially pointing along $x$, as their $x$ position traverses half the system length. To obtain a systematic trend in the presence of rotational noise we average over 50 (12) independent runs in the simulation (experiment) for a given $\phi_b$ and $d_0$. Fig.~\ref{trajectory}b \& d present $\Delta y$ vs bead area fraction $\phi_b$ for two initial separations in experiment and simulation. A systematic trend of a positive $\Delta y$ (effective repulsion) for $\phi_b$ corresponding to the liquid or hexatic for $\phi_b < 0.74$, and negative (effective attraction) for the crystal at $\phi_b>0.74$, is seen, with possible weakening of attraction for $\phi_b = 0.8$ in the simulation. In some cases the rods turn towards each other at the same time after a latency period, but more often one goes straight and the other turns. The lag in the response of the rods is presumably a result of kinetic barriers {associated with the precise arrangement of the beads. These will typically differ for the two rods, so that one of them will get ahead of the other, leading to the observed asymmetry between their trajectories}.

\section{Theory}
Motivated by these observations we construct a theory of the coupled dynamics of $N$ motile rods labelled $\alpha = 1, ..., N$, with positions $\mathbf{R}_{\alpha}(t)$ and orientation unit vectors $\mathbf{n}_{\alpha}(t)$, and a medium characterized by an Eulerian displacement field $\mathbf{u}(\mathbf{x},t)$ as a function of position $\mathbf{x}$ and time $t$, the whole in contact with a structureless substrate that provides a damping linear in velocity, and defines a natural rest frame with respect to which we write our equations of motion. Unlike in \cite{henkes2020dense}, we distinguish the particles driving the medium from those that comprise it. We make several simplifying assumptions and approximations which we list here. We neglect inertia, as the rods don't fly, they walk \cite{narayan2010phase,NKPRL,yamada2003coherent}. We do not resolve the bead-scale structure of the medium and so treat it as translation- and rotation-invariant. We assume that in the absence of rods $\mathbf{u}$ relaxes to minimize an elastic free energy \cite{landau} 
\begin{equation} 
\label{free}
F=\int d^2x\ [\lambda (\mathrm{Tr}\bm{\varepsilon})^2/2 +\mu \mathrm{Tr}(\bm{\varepsilon}^2)]
\end{equation}
with symmetric strain $\bm{\varepsilon}=(\bm{\nabla} \mathbf{u}+\bm{\nabla} \mathbf{u}^ \mathrm{T})/2$ and elastic constants $\lambda$ and $\mu$. We focus on low rod concentrations where the bead medium remains crystalline, and where steric or potential interactions between rods play no role. The absence of interaction between non-motile rods mentioned above assures us that we can ignore elastic couplings between rods as static inclusions \cite{eshelby1957determination}. We focus here on the average behavior but it is straightforward to include noise in our treatment. Within our description each rod simply follows its nose at constant speed $v_0$:  
\begin{equation}
\label{motileEoM} 
\dot{\mathbf{R}}_{\alpha}(t)=v_0 \mathbf{n}_{\alpha}(t),    
\end{equation}
which would describe an active Brownian particle \cite{ebeling2002self,romanczuk2012active,cates2013active} if $\mathbf{n}_{\alpha}$ executed a rotational random walk. The effect of the elastic medium on a motile rod, and of the rods on each other, enters in our theory through rotation of ${\bf n}$ and the resulting re-direction of the velocity of self-propulsion, much as in the imitation of chemotaxis by active colloids \cite{saha2014clusters,zottl2016emergent,liebchen2017phoretic,saha2019pairing,golestanian2019phoretic}.

\begin{figure}[H]
	\begin{center}
		\includegraphics[width=0.45\textwidth]{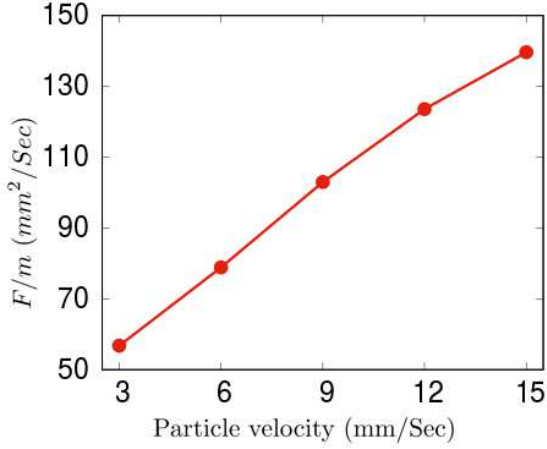}\\
		\caption{Relation between the force F exerted by the motile particle on the elastic bead medium and its velocity $v_0${, as measured in simulations}.}
		\label{rel_fv}
	\end{center}
\end{figure}

As argued in \cite{NatCom,henkes2020dense} the pushing or dragging of the medium with respect to a substrate that serves as a momentum sink endows each motile rod with a monopole force density of strength $f$, consistent with the general properties of swimmers in contact with a substrate \cite{aranson2007model,maitra2019oriented,maitra2020swimmer}:   
\begin{equation}
\label{ueq}
\zeta \partial_t \mathbf{u} = -\frac{\delta F}{\delta \mathbf{u}} + f \sum_{\alpha} \mathbf{n}_{\alpha}(t)\delta(\mathbf{r}-\mathbf{R}_{\alpha}(t)), 
\end{equation}
where $\zeta$ is the coefficient of drag due to the substrate. In Fig.~\ref{rel_fv}, we show that the values of $f$ and $v_0$ inferred from our {simulations} are roughly proportional. {We obtain the relation between self-propelling speed $v_0$ and the force $f$ by simulating polar particles with imposed velocities and measuring the momentum transferred to the beads.}
We proceed by first solving \eqref{ueq} for one motile rod. Let the displacement field in a frame comoving and co-rotating with the rod be $\mathbf{U} = (U_x,U_y)$, which is stationary in an unbounded system if no other rods are present. The lab-frame displacement $\mathbf{u}(\mathbf{r},t)=\mathbf{S}(t)\cdot \mathbf{U}(\mathbf{r}')$, where $\mathbf{S}$ is the rotation matrix from the frame fixed in the rod to the lab frame and $\mathbf{r}'=\mathbf{S}^{\mathrm{T}}\cdot (\mathbf{r}-\mathbf{R}(t))$. {Considering that the motile particle is oriented along the x direction in frame S i.e. $\mathbf{n}(t) = \mathbf{S}(t) \cdot \hat{\mathbf{x}}$. Then, from Eq.~\eqref{ueq}
\begin{eqnarray}\label{ueq1}
\nonumber
\zeta \left[\dfrac{\partial}{\partial t}+\mathbf{S}^{\mathrm{T}}\cdot \dot{\mathbf{S}}\cdot + \mathbf{r}'\cdot\mathbf{S}^{\mathrm{T}} \cdot \dot{\mathbf{S}}\cdot\bm{\nabla}'   \right]\bold{U} = [\zeta v_0 \partial'_x \\+ (\mu \nabla'^2 + \lambda \bm{\nabla}' \bm{\nabla}'\cdot)] \bold{U} + f  \delta(\mathbf{S}\cdot \mathbf{r}') \hat{\mathbf{x}}.
\end{eqnarray}
For the motile particle pointing at angle $\theta(t)$ from the $x$ axis, $\mathbf{n} (t)=(\cos\theta(t),\sin \theta(t))$ and the rotational matrix
\begin{equation}\label{rotmat}
\mathbf{S}=\left( \begin{matrix}
\cos\theta(t) &-\sin \theta(t)\\
\sin \theta(t) & \cos\theta(t)
\end{matrix}\right)
\end{equation}
which gives us  $\mathbf{S}^{\mathrm{T}}\cdot \dot{\mathbf{S}}=\bm{\epsilon}^{\mathrm{T}}\dot{\theta}$. Here $\bm{\epsilon}$ is the 2D Levi-Civita symbol. As $\left| \mathbf{S} \right| =1$, $ \delta(\mathbf{S}\cdot \mathbf{r}')= \delta( \mathbf{r}')$.  Then Eq.~\eqref{ueq1} reduces to
\begin{eqnarray}\label{ueq2}
\nonumber
\zeta \left[\dfrac{\partial \bold{U}}{\partial t}+\left[  \bm{\epsilon}^{\mathrm{T}}\cdot\bold{U} + (\mathbf{r}'\cdot\bm{\epsilon}^{\mathrm{T}} \cdot\bm{\nabla}' )\bold{U}\right] \dot{\theta}(t)   \right] =[\zeta v_0 \partial'_x \\+ (\mu \nabla'^2 + \lambda \bm{\nabla}' \bm{\nabla}'\cdot)] \bold{U} + f  \delta( \mathbf{r}') \hat{\mathbf{x}}.
\end{eqnarray}
For the motile particle subjected to no torque, $\dot{\theta}(t)=0$ and $\bold{U}$ will be constant in time in the stationary state \textit{i.e.} $\bold{U}\equiv\bold{U}(\mathbf{r}')$ and the above equation is simplified to}
\begin{equation}
\label{ueqsimple}
[-\zeta v_0 \partial_x - (\mu \nabla^2 + \lambda \bm{\nabla} \bm{\nabla}\cdot)] \mathbf{U} = f  \delta(\mathbf{r}) \hat{\mathbf{x}},   
\end{equation}
reminiscent of Oseen's modification \cite{oseen1910uber} to the Stokes solution \cite{stokes1850effect} for flow around a slowly moving object, with the difference that \eqref{ueqsimple} describes a displacement {field and not a velocity field}, with the relative importance of advection and diffusive relaxation on a length-scale $L$ encoded not in the Reynolds number but in a P\'{e}clet number $\mbox{Pe}(L) \equiv \zeta v_0 L / \mu$.
The Fourier transform $\mathbf{U}_{\mathbf q} = \int_{\mathbf x} \exp(-i{\mathbf q}\cdot{\mathbf x}) \mathbf{U}({\mathbf x})$ satisfies 
\begin{equation}
- i  v_0 \zeta q_x \mathbf{U}_{\mathbf q} + \mu q^2 \mathbf{U}_{\mathbf q} + \lambda \mathbf{q} \mathbf{q}\cdot\mathbf{U}_{\mathbf q} = f \hat{\mathbf{x}},
\end{equation}
so that 
\begin{eqnarray}
\nonumber
\mathbf{U_q} & =&\frac{f}{\zeta v_0} \left(  \frac{1}{(\xi q^2 - i   q_x)}+ \frac{q_x \xi}{i  (q^2 \xi- i   q_x\nu)}- \frac{q_x\xi}{i (q^2 \xi- i   q_x)}\right) \hat{\mathbf{x}}\\&  +&\left( \frac{i  f q_y \xi}{i  \zeta v_0 (\xi q^2 - i  q_x \nu)} - \frac{i  f q_y\xi}{i  \zeta v_0 (q^2 \xi - i   q_x)}\right) \hat{\mathbf{y}},
\end{eqnarray}
where $\xi=\mu/\zeta v_0$ is a screening length, and $\nu =(1+\eta^{-1})$ with $\eta=\lambda/\mu$. 
Inverting the Fourier transform in terms of tabulated 
integrals~\cite{gradshteyn2014table} yields 
\begin{eqnarray}\label{uxuy}
\nonumber
U_x &=& \frac{f}{4\pi  \mu} \bigg\{ \left[ K_0 \left({r \over 2\xi}\right)-  \frac{x}{r} K_1 \left({r\over 2\xi}\right)\right]e^{-{x}/{2 \xi}}\\
\nonumber
&&+\nu \left[ K_0\left({\nu r\over 2\xi}\right)
+ \frac{x}{r} K_1\left({\nu r\over 2\xi}\right)\right]e^{-{\nu x}/{2\xi}}\bigg\}\\
\nonumber	
U_y&=&\frac{f}{4\pi  \mu}\frac{y}{r} 
\bigg[ \nu  K_1\left({\nu r\over 2\xi}\right)e^{-{\nu x}/{2\xi}}
-  K_1\left({r\over 2\xi}\right)  e^{-{x}/{2\xi}}\bigg]\\
\end{eqnarray} 
for the $x-$ and $y-$components of $\bold{U}$, where $r = \sqrt{x^2 + y^2}$, and $K_0$ and $K_1$ are the modified Bessel functions of the second kind. The force monopole in \eqref{ueqsimple} introduces a quantity $f/\mu$ with units of length \cite{lengthscalefoot}, which gives a natural scale for ${\bf U}$.
\begin{figure}[htb]
	\begin{center}
		\includegraphics[width=0.48\textwidth]{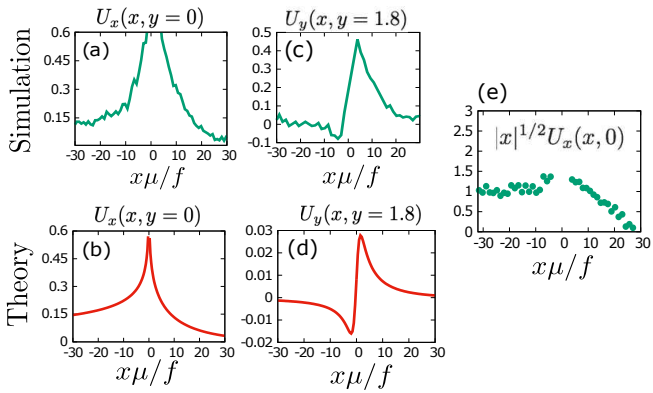}\\
		\caption{Displacement field, scaled by $f/\mu$, around a single motile rod: (a) \& (b) $x$ dependence of $U_x$ at $y=0$ in simulation and theory: both cases see the rapid decay of $U_x(x,y=0)$ for $x>0$ as compared to $x<0$. (c) \& (d) $x$ dependence of $U_y$ at $y\mu/f=1.8$ in simulation and theory. (e) constant unit profile of $|x|^{1/2}U_x(x,0)$ for negative $x$ values shows that $U_x(x,0)$ decays as $1/\sqrt{x}$ behind the particle. For $x>0$, after factoring out $1/\sqrt{x}$, a substantial decay remains, consistent with the predicted exponential but better fitting a linear decrease. In simulation, $\phi_b=0.80$ and $f/ \mu =2.2a $.} \label{ufield}
	\end{center}
\end{figure}

{We extract displacement fields from the distortion of bead-density profiles, focusing here on the single-rod case, using $\rho_0(\mathbf{r}) = \rho(\mathbf{r} +\mathbf{U}(\mathbf{r}))$, where $\rho(\mathbf{r})$ is the time-averaged number density profile in the rest frame of a motile rod, and $\rho_0(\mathbf{r})$ a reference profile obtained by extrapolating from the far field.} {Figs.~\ref{ufield}a \& b show} that the displacement field around a single rod from the simulation at $\phi_b=0.80$ qualitatively matches that predicted by our theory with {$\eta = 1.3$ and $\alpha=0.1$}, except close to the polar rod whose nonzero size is not included in our theory 
The distinctive profile of $U_y$, including its sign change as $x$ goes through zero, are well confirmed by the simulation, as are all qualitative features of the $y$ dependence. 
Crucially, the asymptotic properties of $K_0$ and $K_1$ imply that, for locations along the $x$ axis, the dominant large-$x$ decay of $\mathbf{U}$ is exponential ahead of the motile particle but only as $1/\sqrt{x}$ behind it. The measured displacement field confirms this prediction: Fig.~\ref{ufield}(e) shows that $|x|^{1/2} U_x$ as a function of $x$ for $y$ near $0$ is flat for $x<0$. For $x>0$ by contrast, $|x|^{1/2} U_x$ decays rapidly, though better described by $1 - \mbox{const.}x$ than an exponential.  
We will see below that this extreme asymmetry confers stealth on a motile rod as it approaches another from behind, a feature central to the non-reciprocal pair interaction. 
In next section, we show that the $x$ and $y$ dependence of $U_x$ are described satisfactorily by a common value of the parameters in \eqref{uxuy} (Fig. ~\ref{ufield2}), and discuss possible reasons why measured values of $U_y$ exceed our theoretical estimates.

\subsection{Fitting the numerical displacement field to the theory}
\begin{figure}[htb]
	\begin{center}
		\includegraphics[width=0.5\textwidth]{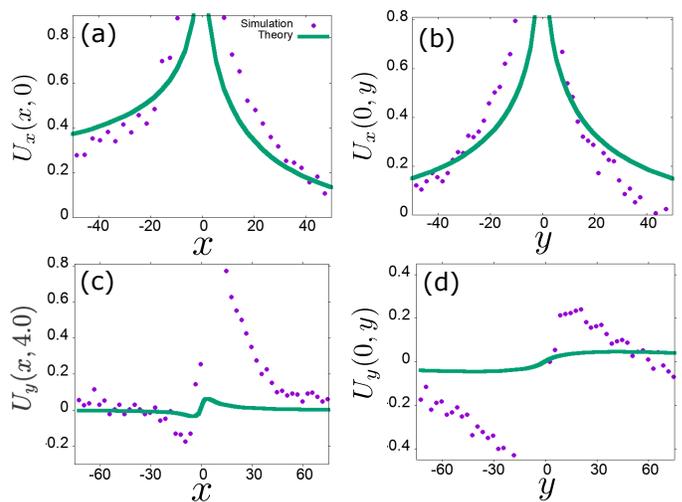}\\
		\caption{The displacement field components $U_x$ and $U_y$ as the function of $x$ and $y$ as observed in the simulation (dots) and corresponding theoretical fits (line). (a) $U_x$ vs $x$ ay $y=0$. (b) $U_x$ vs $y$ at $x=0$. (c) $U_y$ vs $x$ at $y=4a$. (d) $U_y$ vs $y$ at $x=0$. Here the fitting parameters are calculating by fitting the $x$ dependence of $U_x$ at $y=0$ and their values are $\eta=1.3$, $f/\mu=2.2a$ and $\alpha=0.1$. The length unit is bead radius in the simulation.}
		\label{ufield2}
	\end{center}
\end{figure}
\begin{figure}[htb]
	\begin{center}
		\includegraphics[width=0.5\textwidth]{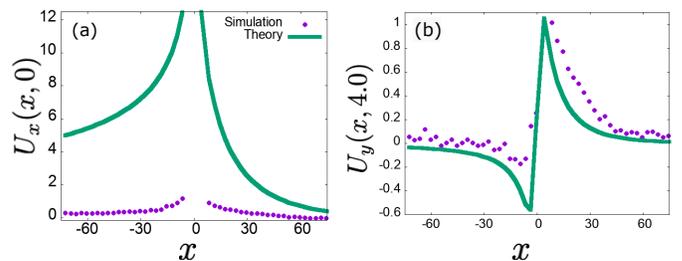}\\
		\caption{{The displacement field components $U_x$ and $U_y$ as the function of $x$ as observed in the simulation (dots) and comparison to theory (line). Here parameters are calculated by fitting the $x$ dependence of $U_y$ at $y=4a$. The resulting values $\eta=1.0$, $f_1/\mu=6.46a$, $f_2 d/\mu=0.10a$ and $\xi=16a$ are then used to generate the curve for $U_x$.} (a) $U_x$ vs $x$ at $y=0$. (b) $U_y$ vs $x$ at $y=4a$.}
		\label{newcal}
	\end{center}
\end{figure}

 
We fit $U_x(x,y=0)$ vs $x$ calculated from the simulation at $\phi_b=0.80$ to our theoretical model (see Eq.~\eqref{uxuy}) using least-squares method (see Fig.~\ref{ufield2}a). The values of fitting parameters are $\eta=1.3$, $f/\mu=2.2a$ and $\alpha=0.1$. Fig.~\ref{ufield2}b shows that these parameter values provide a satisfactory fit to $U_x(x=0,y)$ vs $y$ as well (see Fig.~\ref{ufield2}b). However, the profile of $U_y$ deviates from the theory with these parameter values (see Fig.~\ref{ufield2}c \& d).  {It remains unclear why the $y$ component of the displacement is substantially underestimated by the theory. One possibility is that the motile particle, in addition to exerting a monopole force in the direction in which it points and moves, pushes outward on the medium in the transverse direction. We have attempted to accommodate such a process by extending our calculation to include a force dipole of strength $f_2$ along $y$ in addition to a force monopole of strength $f_1$ along $x$. The result is not encouraging -- unsurprisingly, in retrospect. The relation between forces and elastic displacements is not diagonal, simply because a crystal has a bulk and a shear modulus; a $y$ force dipole large enough to produce a $y$ displacement sufficient to account for the observed value pushes the $x$ displacement to unacceptably large values (Fig. \ref{newcal}). We speculate that the origin of the discrepancy lies outside a linear elastic description, possibly in a dilatant response \cite{reynolds1885lvii} to the shear force exerted by the moving rod on the fluidized medium adjacent to it.}\\


\subsection{Coupled dynamics of two motile particles in an elastic medium  and discussion on coupling terms}
{Let us consider two motile particles{, say, $\alpha$ and $\beta$,} whose orientations and positions at time $t$ are denoted by $\mathbf{n}_{\alpha,\beta}(t)\equiv(\cos\theta_{\alpha,\beta}(t),\sin \theta_{\alpha,\beta}(t))$ and $\mathbf{R}_{\alpha,\beta}(t)$ respectively. Then the displacement field in the lab frame created by $\beta_{th}$ particle will be; $\mathbf{u}_\beta(\mathbf{r},t)=\mathbf{S}_\beta(t)\cdot \mathbf{U}_\beta(\mathbf{r}'_\beta,t)$ with $\mathbf{r}_\beta'(t)=\mathbf{S}_\beta(t)^{\mathrm{T}}\cdot (\mathbf{r}-\mathbf{R}_\beta(t))${, where} the rotation matrix $\mathbf{S}_\beta(t)$ is given by \eqref{rotmat}, and 
the displacement field $\mathbf{U}_\beta(\mathbf{r},t)$ due to motile particle $\beta$ in its co-moving and co-rotating frame follows the equation of motion \eqref{ueq2}.
{Eq. \eqref{motileEoM} tells us that within our treatment each polar rod simply follows its nose. Accordingly, interactions between the rods take place only through changes in their orientations.
The} general form of the interaction of the orientation ${\bf n_{\alpha}}$ with lattice distortions created by particle $\beta$, as permitted by symmetry and at leading orders in gradients, is  given by 
\begin{eqnarray} 
\label{peq}
\dfrac{d\mathbf{n}_{\alpha}}{dt} &=& (\mathbf{I}-\mathbf{n}_{\alpha}\mathbf{n}_{\alpha})\cdot\left[\kappa \bm{\varepsilon_\beta}\cdot \mathbf{n}_{\alpha} + (\gamma \bm{\nabla}\bm{\varepsilon_\beta}) + \ldots\right] 
\end{eqnarray}
where the right-hand side of \eqref{peq} is evaluated at $\mathbf{R}_{\alpha}(t)$. At leading order in gradients \eqref{peq} is invariant under $\mathbf{n}_{\alpha} \to -\mathbf{n}_{\alpha}$: the \textit{apolar} coupling $\kappa$, promoting the alignment of the rod \textit{axis} along a principal direction of $\bm{\varepsilon_\beta}$ is permitted even for a fore-aft symmetric rod \cite{maitra2019oriented}. \textit{Polarity} enters at order $\nabla\bm{\varepsilon_\beta}$, biasing the rod to point in a direction determined by inhomogeneities in strain \cite{banerjee2011substrate,yoshinaga2010polarity}. We denote the relevant coefficients schematically by a single symbol $\gamma$ in \eqref{peq}, though $\nabla\cdot\bm{\varepsilon_\beta}$, $\nabla\mbox{Tr}\bm{\varepsilon_\beta}$, and contractions of $\nabla\bm{\varepsilon_\beta}$ with $\mathbf{n}_{\alpha}$ are all permitted.
The solution \eqref{uxuy} contains a decay length $\xi$, so that $\nabla \bm{\varepsilon_\beta} \sim \xi^{-1} \bm{\varepsilon_\beta}$. 
{The $\gamma$ terms in \eqref{peq} should therefore be of order $\ell/\xi$ relative to the $\kappa$ terms, where $\ell$ is the shape-polarity length defined earlier}. Our neglect of $\gamma$ in \eqref{peq} is thus justified for large $\xi$, i.e., small $v_0$. 

We will therefore focus on the $\kappa$ term on the right of \eqref{peq}, capturing the reorienting effect of distortions to leading order in gradients, but will return briefly to $\gamma$ below. Based on our observations below, we will take $\kappa>0$; i.e., we will consider rods that align with the \textit{extensional} axis of $\bm{\varepsilon}$.
We can easily see from equation (\ref{peq}) that equation of motion for $\theta_\alpha$ will be
\begin{equation}\label{thetaeq}
\dot{\theta}_\alpha(t) = \hat{\mathbf{z}}\cdot[\mathbf{n}_\alpha\times\left(\kappa \bm{\varepsilon}_\beta\cdot \mathbf{n}_\alpha +\gamma \bm{\nabla}\bm{\varepsilon_\beta}	+ ...\right)]
\end{equation}
With the above expression of $\dot{\theta}_\alpha$, the second term 
on the LHS of Eq.\eqref{ueq2} will be a nonlinear term which we ignore in our calculation. Therefore, in the stationary state, the displacement field $\bold{U}_\alpha$ is simply given by Eq.\eqref{uxuy}. From Eq.~\eqref{motileEoM}, the positions $\mathbf{R}_{\alpha,\beta}(t)$ obey
\begin{equation}\label{reqi}
\dot{\mathbf{R}}_{\alpha,\beta}(t)=v_0\mathbf{n}_{\alpha,\beta} (t).
\end{equation}
We integrate Eqs~\eqref{peq} and~\eqref{reqi} numerically to get the trajectories of the two particles in the elastic medium.}

\begin{figure}[htb]
	\begin{center}
		\includegraphics[width=0.48\textwidth]{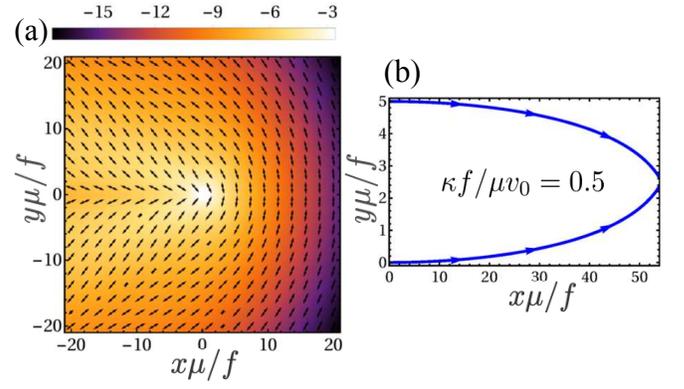}\\
		\caption{{From the displacement field created by a single motile particle moving along the $x$ direction: (a) Principal axis of the traceless part of $\kappa \bm{\varepsilon}$  superimposed with the color map, on a log scale, of its positive eigenvalue at $\kappa f/\mu v_0=1$. (b) Trajectories of the two motile particles coupled with each other's displacement field via Eq.~\eqref{peq} for initial lateral distance $d\mu/f=5$. Here, $\kappa f/\mu v_0=0.5$ 
		$\eta=1$ and $\alpha=1$.}}
		\label{attraction}
	\end{center}
\end{figure} 

Fig. \ref{attraction}a shows the map of the extensional axis of the resulting strain field $\boldsymbol{\varepsilon}$. Through \eqref{peq}, for $\kappa>0$, two such rods, pointing along $\hat{\bf x}$ and separated only along $\hat{\bf y}$, each reorient the other's axis along the field in Fig. \ref{attraction}a and thus locomote towards each other through \eqref{motileEoM}. The resulting trajectories in Fig. \ref{attraction}b account qualitatively for the positive pairwise ``taxis'' of a pair of particles reported in Fig. \ref{trajectory}, modulo the trajectory asymmetry of the latter on which we have already commented above.

\label{gammakappa}
\begin{figure}[H]
	\begin{center}
		\includegraphics[width=0.30\textwidth]{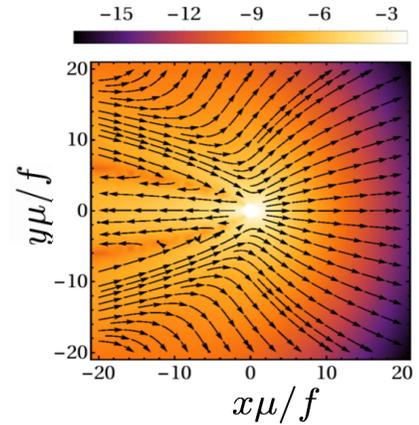}\\
		\caption{{Streamline plot of  $\gamma_1 \nabla^2 {\bf u} + \gamma_2 \nabla \nabla \cdot {\bf u}$, for a single motile particle moving along the x direction, superimposed with the color map of its magnitude. The color map is shown on a log scale.} 
		}
		\label{sup_ufield}
	\end{center}
\end{figure}

\begin{figure}[htb]
	\begin{center}
		\includegraphics[width=0.5\textwidth]{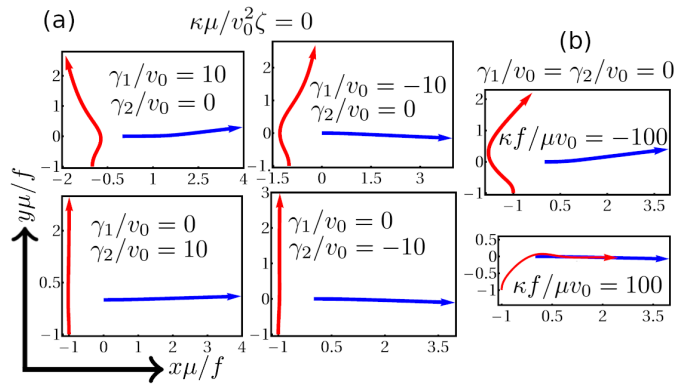}\\
		\caption{Trajectories of two polar particles in theory which are initially pointing normal to each other, (a) for different values of  $\gamma_1/v_0$ and $\gamma_2/v_0$  at $\kappa f/\mu v_0 = 0$, (b)for $\kappa f/\mu v_0 = -100$ and $100$ at $\gamma_1/v_0 = \gamma_2/v_0 = 0$. Here $\alpha = 10$ and $\eta = 1$.	}
		\label{pletraj}
	\end{center}
\end{figure}

We now discuss the polar couplings $\gamma$ in \eqref{peq}, by including terms $\gamma_1\nabla\cdot\bm{\varepsilon}$ and $\gamma_2\nabla\mbox{Tr}\bm{\varepsilon}$. The streamline plots, Fig.~\ref{sup_ufield}, suggest that $\gamma_1,\gamma_2$ give rise to more complex and rapid spatial variation than the apolar strain-aligning $\kappa$ term and do not play an important role in the pair attraction. We illustrate these distinct roles by plotting trajectories calculated by numerically solving \eqref{peq} and \eqref{motileEoM} for the pair, \underline{with $\kappa=0$}, for various positive and negative values of $\gamma_1,\gamma_2$, and see that in all cases capture does not take place. We also show that for $\gamma_1,\gamma_2 = \pm 10$ and $\kappa<0$ again capture is unsuccessful (Fig.~\ref{pletraj}). The reason is that the alignment directions of the rods would now have to be perpendicular to those shown in Fig.~\ref{sup_ufield}. 

The particles in the experiments and simulations are rods tapered towards one tip, and self-propel with the tapered end forward. Considerations of excluded volume suggest that the fat, i.e., non-tapered end would be more easily accommodated in dilated regions, and on the ``outside'' of a curved region of crystal. This suggests that ${\bf n}$ would like to point antiparallel to $\nabla \nabla \cdot {\bf u}$ but parallel to $\nabla^2 {\bf u}$, i.e., $\gamma_1 > 0, \gamma_2 < 0$. However, the interactions could be based on motility in which case essentially entropic excluded-volume arguments are not a good guide. A 
behavioral motility-strain coupling arises in a 
model of 
swarm mechanics \cite{peleg2018collective}, in which bees orient and move towards more dilated regions, 
which would amount to $\gamma_2 > 0$.

\section{Non-reciprocal interactions}
It is clear from the formulation \eqref{peq}-\eqref{ueqsimple} that the interaction between the motile particles in our theory is a consequence of the reorientation of their active motion rather than a pair potential. We now underline the nonequilibrium character of this interaction by a {direct} demonstration of its non-reciprocal nature. Unlike in \cite{fruchart2020phase,ivlev2015statistical,cavagna2017nonsymmetric}, we are dealing here with interactions between two particles of the same type, so non-reciprocality operates with respect to their relative locations and orientations, as in \cite{das2002driven,das2004nonequilibrium,chen2017fore,dadhichi2020nonmutual,saha2019pairing}. The strain field generated by each motile rod reorients the other, thus redirecting their velocities. The fore-aft asymmetry of the lattice distortion in \eqref{uxuy} and Fig. \ref{ufield} implies that if one rod is situated in front of another, as defined by the heading of the latter, the two experience drastically different reorienting torques. The strain field generated ahead of the trailing rod is highly screened, unlike that which the leading rod produces in its wake. We expect the trailing rod to reorient strongly, while effectively concealing its approach from the leading rod. We illustrate this through theoretically calculated particle trajectories which we compare with {laboratory and} numerical experiments.
	
\begin{figure}[htb]
	\begin{center}
		\includegraphics[width=0.48\textwidth]{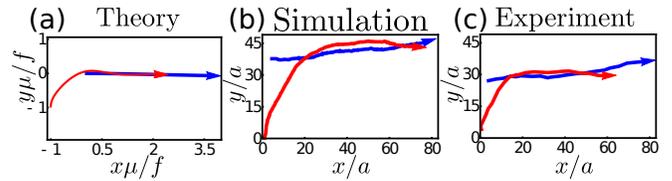}\\
		\caption{Nonreciprocality in capture. Trajectories of two polar particles which are initially pointing normal to each other. (a) in theory for 
			$\kappa f / \mu v_0 = 100$, {(b) in simulation and (c) in experiment at $\phi_b=0.78$ for initial $y$ separation 39.4$a$ and 26$a$ respectively.}}
		\label{nonreciptheory}
	\end{center}
\end{figure}

We consider a pair of particles initially aligned perpendicular to each other, pointing along say the $x$ and $y$ directions. 
Eq. \eqref{peq} with $\gamma = 0$ and $\kappa > 0$ always leads to capture, as illustrated in Fig. \ref{nonreciptheory}a for $\kappa f /\mu v_0 = 100$. The calculated dynamics {graphically} reiterates what we argued qualitatively above: the encounter takes place by one particle approaching the other from behind, and turning towards it, while the particle in front moves entirely oblivious of the approach of the other, see Fig.~\ref{nonreciptheory}a, a direct consequence of the strong fore-aft asymmetry in the screening of the displacement fields of the moving particles, \eqref{uxuy}. We find {convincing} confirmation of this prediction in our simulation as well as in experiment, see Fig. \ref{nonreciptheory}b \& c and Supplementary Movie S9, S10 \cite{supp}.

\section{Conclusion}
We have thus shown, through experiments and simulations on vibrated grains, that active particles can locomote through a supported two-dimensional crystalline medium, which mediates an effective attraction between them by reorienting their direction of motion. We have accounted for our observations through a theory of moving force monopoles, which implies a non-potential sensing and trail-following behaviour of purely mechanical origin. Our measurements confirm {qualitatively} the predicted form of the strain field of a single motile particle and the non-reciprocal attraction and pursuit of a pair. This interaction should be observed between motile cells in a supported gel layer or epithelium, and possibly in chase-and-run dynamics in cell migration \cite{theveneau2013chase,mayett2017chase}. {Despite the qualitative success of the theory, we find it substantially underestimates the component of the displacement field of the medium transverse to (or overestimates the component parallel to) the direction of motion of the active particle. We speculate that the explanation may lie in dilatancy.} Modifications associated with momentum conservation arise in principle for a bulk 3-dimensional medium such as the extracellular matrix but even there the relatively rigid components of the matrix could serve, over some time- and length-scales, as a momentum sink for the dynamics of the more deformable components. Multiparticle behaviour, the competition between alignment and attraction in the collective dynamics, and the effect of quenched disorder, as well as the relation between the parameters of our coarse-grained theory and particle-scale properties remain outstanding challenges.

\begin{acknowledgments}
RKG thanks the Department of Physics, Indian Institute of Science, for hospitality and the SERB, India, for support. RK was supported by the University Grants Commission, AKS by a Year of Science Professorship from the DST, HS by SERB, India and SR by a J C Bose Fellowship from SERB and by the Tata Education \& Development Trust. SR thanks E Tjhung and M Cates for discussions, and DAMTP, Cambridge for support and hospitality, funded in part by the European Research Council under the EU’s Horizon 2020 Programme, Grant number 740269. SR and AKS acknowledge illuminating discussions during the KITP 2020 online program on Symmetry, Thermodynamics and Topology in Active Matter, {supported by} the National Science Foundation under Grant No. NSF PHY-1748958.
\end{acknowledgments}

\bibliography{ref}

\begin{thebibliography}{71}%
\makeatletter
\providecommand \@ifxundefined [1]{%
 \@ifx{#1\undefined}
}%
\providecommand \@ifnum [1]{%
 \ifnum #1\expandafter \@firstoftwo
 \else \expandafter \@secondoftwo
 \fi
}%
\providecommand \@ifx [1]{%
 \ifx #1\expandafter \@firstoftwo
 \else \expandafter \@secondoftwo
 \fi
}%
\providecommand \natexlab [1]{#1}%
\providecommand \enquote  [1]{``#1''}%
\providecommand \bibnamefont  [1]{#1}%
\providecommand \bibfnamefont [1]{#1}%
\providecommand \citenamefont [1]{#1}%
\providecommand \href@noop [0]{\@secondoftwo}%
\providecommand \href [0]{\begingroup \@sanitize@url \@href}%
\providecommand \@href[1]{\@@startlink{#1}\@@href}%
\providecommand \@@href[1]{\endgroup#1\@@endlink}%
\providecommand \@sanitize@url [0]{\catcode `\\12\catcode `\$12\catcode
  `\&12\catcode `\#12\catcode `\^12\catcode `\_12\catcode `\%12\relax}%
\providecommand \@@startlink[1]{}%
\providecommand \@@endlink[0]{}%
\providecommand \url  [0]{\begingroup\@sanitize@url \@url }%
\providecommand \@url [1]{\endgroup\@href {#1}{\urlprefix }}%
\providecommand \urlprefix  [0]{URL }%
\providecommand \Eprint [0]{\href }%
\providecommand \doibase [0]{http://dx.doi.org/}%
\providecommand \selectlanguage [0]{\@gobble}%
\providecommand \bibinfo  [0]{\@secondoftwo}%
\providecommand \bibfield  [0]{\@secondoftwo}%
\providecommand \translation [1]{[#1]}%
\providecommand \BibitemOpen [0]{}%
\providecommand \bibitemStop [0]{}%
\providecommand \bibitemNoStop [0]{.\EOS\space}%
\providecommand \EOS [0]{\spacefactor3000\relax}%
\providecommand \BibitemShut  [1]{\csname bibitem#1\endcsname}%
\let\auto@bib@innerbib\@empty
\bibitem [{\citenamefont {Pooley}\ \emph {et~al.}(2007)\citenamefont {Pooley},
  \citenamefont {Alexander},\ and\ \citenamefont {Yeomans}}]{YeomansPrl2007}%
  \BibitemOpen
  \bibfield  {author} {\bibinfo {author} {\bibfnamefont {C.~M.}\ \bibnamefont
  {Pooley}}, \bibinfo {author} {\bibfnamefont {G.~P.}\ \bibnamefont
  {Alexander}}, \ and\ \bibinfo {author} {\bibfnamefont {J.~M.}\ \bibnamefont
  {Yeomans}},\ }\href {\doibase 10.1103/PhysRevLett.99.228103} {\bibfield
  {journal} {\bibinfo  {journal} {Phys. Rev. Lett.}\ }\textbf {\bibinfo
  {volume} {99}},\ \bibinfo {pages} {228103} (\bibinfo {year}
  {2007})}\BibitemShut {NoStop}%
\bibitem [{\citenamefont {Baskaran}\ and\ \citenamefont
  {Marchetti}(2009)}]{baskaran2009statistical}%
  \BibitemOpen
  \bibfield  {author} {\bibinfo {author} {\bibfnamefont {A.}~\bibnamefont
  {Baskaran}}\ and\ \bibinfo {author} {\bibfnamefont {M.~C.}\ \bibnamefont
  {Marchetti}},\ }\href {https://doi.org/10.1073/pnas.0906586106} {\bibfield
  {journal} {\bibinfo  {journal} {Proceedings of the National Academy of
  Sciences}\ }\textbf {\bibinfo {volume} {106}},\ \bibinfo {pages} {15567}
  (\bibinfo {year} {2009})}\BibitemShut {NoStop}%
\bibitem [{\citenamefont {Simha}\ and\ \citenamefont
  {Ramaswamy}(2002)}]{simha2002hydrodynamic}%
  \BibitemOpen
  \bibfield  {author} {\bibinfo {author} {\bibfnamefont {R.~A.}\ \bibnamefont
  {Simha}}\ and\ \bibinfo {author} {\bibfnamefont {S.}~\bibnamefont
  {Ramaswamy}},\ }\href {https://doi.org/10.1103/PhysRevLett.89.058101}
  {\bibfield  {journal} {\bibinfo  {journal} {Physical Review Letters}\
  }\textbf {\bibinfo {volume} {89}},\ \bibinfo {pages} {058101} (\bibinfo
  {year} {2002})}\BibitemShut {NoStop}%
\bibitem [{\citenamefont {Marchetti}\ \emph {et~al.}(2013)\citenamefont
  {Marchetti}, \citenamefont {Joanny}, \citenamefont {Ramaswamy}, \citenamefont
  {Liverpool}, \citenamefont {Prost}, \citenamefont {Rao},\ and\ \citenamefont
  {Simha}}]{SriramRMP}%
  \BibitemOpen
  \bibfield  {author} {\bibinfo {author} {\bibfnamefont {M.~C.}\ \bibnamefont
  {Marchetti}}, \bibinfo {author} {\bibfnamefont {J.~F.}\ \bibnamefont
  {Joanny}}, \bibinfo {author} {\bibfnamefont {S.}~\bibnamefont {Ramaswamy}},
  \bibinfo {author} {\bibfnamefont {T.~B.}\ \bibnamefont {Liverpool}}, \bibinfo
  {author} {\bibfnamefont {J.}~\bibnamefont {Prost}}, \bibinfo {author}
  {\bibfnamefont {M.}~\bibnamefont {Rao}}, \ and\ \bibinfo {author}
  {\bibfnamefont {R.~A.}\ \bibnamefont {Simha}},\ }\href {\doibase
  10.1103/RevModPhys.85.1143} {\bibfield  {journal} {\bibinfo  {journal} {Rev.
  Mod. Phys.}\ }\textbf {\bibinfo {volume} {85}},\ \bibinfo {pages} {1143}
  (\bibinfo {year} {2013})}\BibitemShut {NoStop}%
\bibitem [{\citenamefont {Saintillan}\ and\ \citenamefont
  {Shelley}(2015)}]{saintillan2015theory}%
  \BibitemOpen
  \bibfield  {author} {\bibinfo {author} {\bibfnamefont {D.}~\bibnamefont
  {Saintillan}}\ and\ \bibinfo {author} {\bibfnamefont {M.~J.}\ \bibnamefont
  {Shelley}},\ }in\ \href {https://doi.org/10.1007/978-1-4939-2065-5_9} {\emph
  {\bibinfo {booktitle} {Complex Fluids in biological systems}}}\ (\bibinfo
  {publisher} {Springer},\ \bibinfo {year} {2015})\ pp.\ \bibinfo {pages}
  {319--355}\BibitemShut {NoStop}%
\bibitem [{\citenamefont {Elgeti}\ \emph {et~al.}(2015)\citenamefont {Elgeti},
  \citenamefont {Winkler},\ and\ \citenamefont {Gompper}}]{elgeti2015physics}%
  \BibitemOpen
  \bibfield  {author} {\bibinfo {author} {\bibfnamefont {J.}~\bibnamefont
  {Elgeti}}, \bibinfo {author} {\bibfnamefont {R.~G.}\ \bibnamefont {Winkler}},
  \ and\ \bibinfo {author} {\bibfnamefont {G.}~\bibnamefont {Gompper}},\ }\href
  {https://doi.org/10.1088/0034-4885/78/5/056601} {\bibfield  {journal}
  {\bibinfo  {journal} {Reports on progress in physics}\ }\textbf {\bibinfo
  {volume} {78}},\ \bibinfo {pages} {056601} (\bibinfo {year}
  {2015})}\BibitemShut {NoStop}%
\bibitem [{\citenamefont {Golestanian}(2019)}]{golestanian2019phoretic}%
  \BibitemOpen
  \bibfield  {author} {\bibinfo {author} {\bibfnamefont {R.}~\bibnamefont
  {Golestanian}},\ }\href {http://arxiv.org/abs/1909.03747} {\bibfield
  {journal} {\bibinfo  {journal} {arXiv preprint arXiv:1909.03747}\ } (\bibinfo
  {year} {2019})}\BibitemShut {NoStop}%
\bibitem [{\citenamefont {Granek}\ \emph {et~al.}(2020)\citenamefont {Granek},
  \citenamefont {Baek}, \citenamefont {Kafri},\ and\ \citenamefont
  {Solon}}]{granek2020bodies}%
  \BibitemOpen
  \bibfield  {author} {\bibinfo {author} {\bibfnamefont {O.}~\bibnamefont
  {Granek}}, \bibinfo {author} {\bibfnamefont {Y.}~\bibnamefont {Baek}},
  \bibinfo {author} {\bibfnamefont {Y.}~\bibnamefont {Kafri}}, \ and\ \bibinfo
  {author} {\bibfnamefont {A.~P.}\ \bibnamefont {Solon}},\ }\href {\doibase
  10.1088/1742-5468/ab7f34} {\bibfield  {journal} {\bibinfo  {journal} {Journal
  of Statistical Mechanics: Theory and Experiment}\ }\textbf {\bibinfo {volume}
  {2020}},\ \bibinfo {pages} {063211} (\bibinfo {year} {2020})}\BibitemShut
  {NoStop}%
\bibitem [{\citenamefont {Bartolo}\ \emph {et~al.}(2003)\citenamefont
  {Bartolo}, \citenamefont {Ajdari},\ and\ \citenamefont
  {Fournier}}]{bartolo2003effective}%
  \BibitemOpen
  \bibfield  {author} {\bibinfo {author} {\bibfnamefont {D.}~\bibnamefont
  {Bartolo}}, \bibinfo {author} {\bibfnamefont {A.}~\bibnamefont {Ajdari}}, \
  and\ \bibinfo {author} {\bibfnamefont {J.-B.}\ \bibnamefont {Fournier}},\
  }\href {https://doi.org/10.1103/PhysRevE.67.061112} {\bibfield  {journal}
  {\bibinfo  {journal} {Physical Review E}\ }\textbf {\bibinfo {volume} {67}},\
  \bibinfo {pages} {061112} (\bibinfo {year} {2003})}\BibitemShut {NoStop}%
\bibitem [{\citenamefont {Schwarz}\ and\ \citenamefont
  {Safran}(2002)}]{SafranPrl2002}%
  \BibitemOpen
  \bibfield  {author} {\bibinfo {author} {\bibfnamefont {U.~S.}\ \bibnamefont
  {Schwarz}}\ and\ \bibinfo {author} {\bibfnamefont {S.~A.}\ \bibnamefont
  {Safran}},\ }\href {\doibase 10.1103/PhysRevLett.88.048102} {\bibfield
  {journal} {\bibinfo  {journal} {Phys. Rev. Lett.}\ }\textbf {\bibinfo
  {volume} {88}},\ \bibinfo {pages} {048102} (\bibinfo {year}
  {2002})}\BibitemShut {NoStop}%
\bibitem [{\citenamefont {Bischofs}\ \emph {et~al.}(2004)\citenamefont
  {Bischofs}, \citenamefont {Safran},\ and\ \citenamefont
  {Schwarz}}]{SafranPre2004}%
  \BibitemOpen
  \bibfield  {author} {\bibinfo {author} {\bibfnamefont {I.~B.}\ \bibnamefont
  {Bischofs}}, \bibinfo {author} {\bibfnamefont {S.~A.}\ \bibnamefont
  {Safran}}, \ and\ \bibinfo {author} {\bibfnamefont {U.~S.}\ \bibnamefont
  {Schwarz}},\ }\href {\doibase 10.1103/PhysRevE.69.021911} {\bibfield
  {journal} {\bibinfo  {journal} {Phys. Rev. E}\ }\textbf {\bibinfo {volume}
  {69}},\ \bibinfo {pages} {021911} (\bibinfo {year} {2004})}\BibitemShut
  {NoStop}%
\bibitem [{\citenamefont {Yuval}\ and\ \citenamefont
  {Safran}(2013)}]{SafranPre2013}%
  \BibitemOpen
  \bibfield  {author} {\bibinfo {author} {\bibfnamefont {J.}~\bibnamefont
  {Yuval}}\ and\ \bibinfo {author} {\bibfnamefont {S.~A.}\ \bibnamefont
  {Safran}},\ }\href {\doibase 10.1103/PhysRevE.87.042703} {\bibfield
  {journal} {\bibinfo  {journal} {Phys. Rev. E}\ }\textbf {\bibinfo {volume}
  {87}},\ \bibinfo {pages} {042703} (\bibinfo {year} {2013})}\BibitemShut
  {NoStop}%
\bibitem [{\citenamefont {Cohen}\ and\ \citenamefont
  {Safran}(2016)}]{safranSoftM2016}%
  \BibitemOpen
  \bibfield  {author} {\bibinfo {author} {\bibfnamefont {O.}~\bibnamefont
  {Cohen}}\ and\ \bibinfo {author} {\bibfnamefont {S.~A.}\ \bibnamefont
  {Safran}},\ }\href {\doibase 10.1039/C6SM00351F} {\bibfield  {journal}
  {\bibinfo  {journal} {Soft Matter}\ }\textbf {\bibinfo {volume} {12}},\
  \bibinfo {pages} {6088} (\bibinfo {year} {2016})}\BibitemShut {NoStop}%
\bibitem [{\citenamefont {Golkov}\ and\ \citenamefont
  {Shokef}(2017)}]{golkov2017shape}%
  \BibitemOpen
  \bibfield  {author} {\bibinfo {author} {\bibfnamefont {R.}~\bibnamefont
  {Golkov}}\ and\ \bibinfo {author} {\bibfnamefont {Y.}~\bibnamefont
  {Shokef}},\ }\href
  {https://iopscience.iop.org/article/10.1088/1367-2630/aa70ef/meta} {\bibfield
   {journal} {\bibinfo  {journal} {New Journal of Physics}\ }\textbf {\bibinfo
  {volume} {19}},\ \bibinfo {pages} {063011} (\bibinfo {year}
  {2017})}\BibitemShut {NoStop}%
\bibitem [{\citenamefont {Golkov}\ and\ \citenamefont
  {Shokef}(2019)}]{golkov2019elastic}%
  \BibitemOpen
  \bibfield  {author} {\bibinfo {author} {\bibfnamefont {R.}~\bibnamefont
  {Golkov}}\ and\ \bibinfo {author} {\bibfnamefont {Y.}~\bibnamefont
  {Shokef}},\ }\href
  {https://journals.aps.org/pre/abstract/10.1103/PhysRevE.99.032418} {\bibfield
   {journal} {\bibinfo  {journal} {Physical Review E}\ }\textbf {\bibinfo
  {volume} {99}},\ \bibinfo {pages} {032418} (\bibinfo {year}
  {2019})}\BibitemShut {NoStop}%
\bibitem [{\citenamefont {Marcq}(2014)}]{marcq2014spatio}%
  \BibitemOpen
  \bibfield  {author} {\bibinfo {author} {\bibfnamefont {P.}~\bibnamefont
  {Marcq}},\ }\href {https://doi.org/10.1140/epje/i2014-14029-x} {\bibfield
  {journal} {\bibinfo  {journal} {The European Physical Journal E}\ }\textbf
  {\bibinfo {volume} {37}},\ \bibinfo {pages} {29} (\bibinfo {year}
  {2014})}\BibitemShut {NoStop}%
\bibitem [{\citenamefont {Banerjee}\ and\ \citenamefont
  {Marchetti}(2011)}]{banerjee2011substrate}%
  \BibitemOpen
  \bibfield  {author} {\bibinfo {author} {\bibfnamefont {S.}~\bibnamefont
  {Banerjee}}\ and\ \bibinfo {author} {\bibfnamefont {M.~C.}\ \bibnamefont
  {Marchetti}},\ }\href {https://doi.org/10.1209/0295-5075/96/28003} {\bibfield
   {journal} {\bibinfo  {journal} {EPL (Europhysics Letters)}\ }\textbf
  {\bibinfo {volume} {96}},\ \bibinfo {pages} {28003} (\bibinfo {year}
  {2011})}\BibitemShut {NoStop}%
\bibitem [{\citenamefont {Banerjee}\ \emph {et~al.}(2015)\citenamefont
  {Banerjee}, \citenamefont {Utuje},\ and\ \citenamefont
  {Marchetti}}]{banerjee2015propagating}%
  \BibitemOpen
  \bibfield  {author} {\bibinfo {author} {\bibfnamefont {S.}~\bibnamefont
  {Banerjee}}, \bibinfo {author} {\bibfnamefont {K.~J.}\ \bibnamefont {Utuje}},
  \ and\ \bibinfo {author} {\bibfnamefont {M.~C.}\ \bibnamefont {Marchetti}},\
  }\href {https://doi.org/10.1103/PhysRevLett.114.228101} {\bibfield  {journal}
  {\bibinfo  {journal} {Physical Review Letters}\ }\textbf {\bibinfo {volume}
  {114}},\ \bibinfo {pages} {228101} (\bibinfo {year} {2015})}\BibitemShut
  {NoStop}%
\bibitem [{\citenamefont {K{\"o}pf}\ and\ \citenamefont
  {Pismen}(2013)}]{kopf2013non}%
  \BibitemOpen
  \bibfield  {author} {\bibinfo {author} {\bibfnamefont {M.}~\bibnamefont
  {K{\"o}pf}}\ and\ \bibinfo {author} {\bibfnamefont {L.}~\bibnamefont
  {Pismen}},\ }\href {https://doi.org/10.1016/j.physd.2013.05.009} {\bibfield
  {journal} {\bibinfo  {journal} {Physica D: Nonlinear Phenomena}\ }\textbf
  {\bibinfo {volume} {259}},\ \bibinfo {pages} {48} (\bibinfo {year}
  {2013})}\BibitemShut {NoStop}%
\bibitem [{\citenamefont {Notbohm}\ \emph {et~al.}(2016)\citenamefont
  {Notbohm}, \citenamefont {Banerjee}, \citenamefont {Utuje}, \citenamefont
  {Gweon}, \citenamefont {Jang}, \citenamefont {Park}, \citenamefont {Shin},
  \citenamefont {Butler}, \citenamefont {Fredberg},\ and\ \citenamefont
  {Marchetti}}]{notbohm2016cellular}%
  \BibitemOpen
  \bibfield  {author} {\bibinfo {author} {\bibfnamefont {J.}~\bibnamefont
  {Notbohm}}, \bibinfo {author} {\bibfnamefont {S.}~\bibnamefont {Banerjee}},
  \bibinfo {author} {\bibfnamefont {K.~J.}\ \bibnamefont {Utuje}}, \bibinfo
  {author} {\bibfnamefont {B.}~\bibnamefont {Gweon}}, \bibinfo {author}
  {\bibfnamefont {H.}~\bibnamefont {Jang}}, \bibinfo {author} {\bibfnamefont
  {Y.}~\bibnamefont {Park}}, \bibinfo {author} {\bibfnamefont {J.}~\bibnamefont
  {Shin}}, \bibinfo {author} {\bibfnamefont {J.~P.}\ \bibnamefont {Butler}},
  \bibinfo {author} {\bibfnamefont {J.~J.}\ \bibnamefont {Fredberg}}, \ and\
  \bibinfo {author} {\bibfnamefont {M.~C.}\ \bibnamefont {Marchetti}},\ }\href
  {https://doi.org/10.1016/j.bpj.2016.05.019} {\bibfield  {journal} {\bibinfo
  {journal} {Biophysical Journal}\ }\textbf {\bibinfo {volume} {110}},\
  \bibinfo {pages} {2729} (\bibinfo {year} {2016})}\BibitemShut {NoStop}%
\bibitem [{\citenamefont {Alaimo}\ \emph {et~al.}(2016)\citenamefont {Alaimo},
  \citenamefont {Praetorius},\ and\ \citenamefont
  {Voigt}}]{alaimo2016microscopic}%
  \BibitemOpen
  \bibfield  {author} {\bibinfo {author} {\bibfnamefont {F.}~\bibnamefont
  {Alaimo}}, \bibinfo {author} {\bibfnamefont {S.}~\bibnamefont {Praetorius}},
  \ and\ \bibinfo {author} {\bibfnamefont {A.}~\bibnamefont {Voigt}},\ }\href
  {https://doi.org/10.1088/1367-2630/18/8/083008} {\bibfield  {journal}
  {\bibinfo  {journal} {New Journal of Physics}\ }\textbf {\bibinfo {volume}
  {18}},\ \bibinfo {pages} {083008} (\bibinfo {year} {2016})}\BibitemShut
  {NoStop}%
\bibitem [{\citenamefont {Menzel}\ and\ \citenamefont
  {L{\"o}wen}(2013)}]{menzel2013traveling}%
  \BibitemOpen
  \bibfield  {author} {\bibinfo {author} {\bibfnamefont {A.~M.}\ \bibnamefont
  {Menzel}}\ and\ \bibinfo {author} {\bibfnamefont {H.}~\bibnamefont
  {L{\"o}wen}},\ }\href {https://doi.org/10.1103/PhysRevLett.110.055702}
  {\bibfield  {journal} {\bibinfo  {journal} {Physical Review Letters}\
  }\textbf {\bibinfo {volume} {110}},\ \bibinfo {pages} {055702} (\bibinfo
  {year} {2013})}\BibitemShut {NoStop}%
\bibitem [{\citenamefont {Maitra}\ and\ \citenamefont
  {Ramaswamy}(2019)}]{maitra2019oriented}%
  \BibitemOpen
  \bibfield  {author} {\bibinfo {author} {\bibfnamefont {A.}~\bibnamefont
  {Maitra}}\ and\ \bibinfo {author} {\bibfnamefont {S.}~\bibnamefont
  {Ramaswamy}},\ }\href {https://doi.org/10.1103/PhysRevLett.123.238001}
  {\bibfield  {journal} {\bibinfo  {journal} {Physical Review Letters}\
  }\textbf {\bibinfo {volume} {123}},\ \bibinfo {pages} {238001} (\bibinfo
  {year} {2019})}\BibitemShut {NoStop}%
\bibitem [{\citenamefont {Henkes}\ \emph {et~al.}(2020)\citenamefont {Henkes},
  \citenamefont {Kostanjevec}, \citenamefont {Collinson}, \citenamefont
  {Sknepnek},\ and\ \citenamefont {Bertin}}]{henkes2020dense}%
  \BibitemOpen
  \bibfield  {author} {\bibinfo {author} {\bibfnamefont {S.}~\bibnamefont
  {Henkes}}, \bibinfo {author} {\bibfnamefont {K.}~\bibnamefont {Kostanjevec}},
  \bibinfo {author} {\bibfnamefont {J.~M.}\ \bibnamefont {Collinson}}, \bibinfo
  {author} {\bibfnamefont {R.}~\bibnamefont {Sknepnek}}, \ and\ \bibinfo
  {author} {\bibfnamefont {E.}~\bibnamefont {Bertin}},\ }\href
  {https://doi.org/10.1038/s41467-020-15164-5} {\bibfield  {journal} {\bibinfo
  {journal} {Nature communications}\ }\textbf {\bibinfo {volume} {11}},\
  \bibinfo {pages} {1405} (\bibinfo {year} {2020})}\BibitemShut {NoStop}%
\bibitem [{\citenamefont {Das}\ \emph {et~al.}(2002)\citenamefont {Das},
  \citenamefont {Rao},\ and\ \citenamefont {Ramaswamy}}]{das2002driven}%
  \BibitemOpen
  \bibfield  {author} {\bibinfo {author} {\bibfnamefont {J.}~\bibnamefont
  {Das}}, \bibinfo {author} {\bibfnamefont {M.}~\bibnamefont {Rao}}, \ and\
  \bibinfo {author} {\bibfnamefont {S.}~\bibnamefont {Ramaswamy}},\ }\href
  {https://doi.org/10.1209/epl/i2002-00280-2} {\bibfield  {journal} {\bibinfo
  {journal} {EPL (Europhysics Letters)}\ }\textbf {\bibinfo {volume} {60}},\
  \bibinfo {pages} {418} (\bibinfo {year} {2002})}\BibitemShut {NoStop}%
\bibitem [{\citenamefont {Das}\ \emph {et~al.}(2004)\citenamefont {Das},
  \citenamefont {Rao},\ and\ \citenamefont
  {Ramaswamy}}]{das2004nonequilibrium}%
  \BibitemOpen
  \bibfield  {author} {\bibinfo {author} {\bibfnamefont {J.}~\bibnamefont
  {Das}}, \bibinfo {author} {\bibfnamefont {M.}~\bibnamefont {Rao}}, \ and\
  \bibinfo {author} {\bibfnamefont {S.}~\bibnamefont {Ramaswamy}},\ }\href
  {https://arxiv.org/abs/cond-mat/0404071} {\bibfield  {journal} {\bibinfo
  {journal} {arXiv:cond-mat/0404071}\ } (\bibinfo {year} {2004})}\BibitemShut
  {NoStop}%
\bibitem [{\citenamefont {Ivlev}\ \emph {et~al.}(2015)\citenamefont {Ivlev},
  \citenamefont {Bartnick}, \citenamefont {Heinen}, \citenamefont {Du},
  \citenamefont {Nosenko},\ and\ \citenamefont
  {L{\"o}wen}}]{ivlev2015statistical}%
  \BibitemOpen
  \bibfield  {author} {\bibinfo {author} {\bibfnamefont {A.}~\bibnamefont
  {Ivlev}}, \bibinfo {author} {\bibfnamefont {J.}~\bibnamefont {Bartnick}},
  \bibinfo {author} {\bibfnamefont {M.}~\bibnamefont {Heinen}}, \bibinfo
  {author} {\bibfnamefont {C.-R.}\ \bibnamefont {Du}}, \bibinfo {author}
  {\bibfnamefont {V.}~\bibnamefont {Nosenko}}, \ and\ \bibinfo {author}
  {\bibfnamefont {H.}~\bibnamefont {L{\"o}wen}},\ }\href
  {https://journals.aps.org/prx/abstract/10.1103/PhysRevX.5.011035} {\bibfield
  {journal} {\bibinfo  {journal} {Physical Review X}\ }\textbf {\bibinfo
  {volume} {5}},\ \bibinfo {pages} {011035} (\bibinfo {year}
  {2015})}\BibitemShut {NoStop}%
\bibitem [{\citenamefont {Cavagna}\ \emph {et~al.}(2017)\citenamefont
  {Cavagna}, \citenamefont {Giardina}, \citenamefont {Jelic}, \citenamefont
  {Melillo}, \citenamefont {Parisi}, \citenamefont {Silvestri},\ and\
  \citenamefont {Viale}}]{cavagna2017nonsymmetric}%
  \BibitemOpen
  \bibfield  {author} {\bibinfo {author} {\bibfnamefont {A.}~\bibnamefont
  {Cavagna}}, \bibinfo {author} {\bibfnamefont {I.}~\bibnamefont {Giardina}},
  \bibinfo {author} {\bibfnamefont {A.}~\bibnamefont {Jelic}}, \bibinfo
  {author} {\bibfnamefont {S.}~\bibnamefont {Melillo}}, \bibinfo {author}
  {\bibfnamefont {L.}~\bibnamefont {Parisi}}, \bibinfo {author} {\bibfnamefont
  {E.}~\bibnamefont {Silvestri}}, \ and\ \bibinfo {author} {\bibfnamefont
  {M.}~\bibnamefont {Viale}},\ }\href
  {https://journals.aps.org/prl/abstract/10.1103/PhysRevLett.118.138003}
  {\bibfield  {journal} {\bibinfo  {journal} {Physical Review Letters}\
  }\textbf {\bibinfo {volume} {118}},\ \bibinfo {pages} {138003} (\bibinfo
  {year} {2017})}\BibitemShut {NoStop}%
\bibitem [{\citenamefont {Chen}\ \emph {et~al.}(2017)\citenamefont {Chen},
  \citenamefont {Patelli}, \citenamefont {Chat{\'e}}, \citenamefont {Ma},\ and\
  \citenamefont {Shi}}]{chen2017fore}%
  \BibitemOpen
  \bibfield  {author} {\bibinfo {author} {\bibfnamefont {Q.-s.}\ \bibnamefont
  {Chen}}, \bibinfo {author} {\bibfnamefont {A.}~\bibnamefont {Patelli}},
  \bibinfo {author} {\bibfnamefont {H.}~\bibnamefont {Chat{\'e}}}, \bibinfo
  {author} {\bibfnamefont {Y.-q.}\ \bibnamefont {Ma}}, \ and\ \bibinfo {author}
  {\bibfnamefont {X.-q.}\ \bibnamefont {Shi}},\ }\href
  {https://journals.aps.org/pre/abstract/10.1103/PhysRevE.96.020601} {\bibfield
   {journal} {\bibinfo  {journal} {Physical Review E}\ }\textbf {\bibinfo
  {volume} {96}},\ \bibinfo {pages} {020601(R)} (\bibinfo {year}
  {2017})}\BibitemShut {NoStop}%
\bibitem [{\citenamefont {Saha}\ \emph {et~al.}(2019)\citenamefont {Saha},
  \citenamefont {Ramaswamy},\ and\ \citenamefont
  {Golestanian}}]{saha2019pairing}%
  \BibitemOpen
  \bibfield  {author} {\bibinfo {author} {\bibfnamefont {S.}~\bibnamefont
  {Saha}}, \bibinfo {author} {\bibfnamefont {S.}~\bibnamefont {Ramaswamy}}, \
  and\ \bibinfo {author} {\bibfnamefont {R.}~\bibnamefont {Golestanian}},\
  }\href {https://doi.org/10.1088/1367-2630/ab20fd} {\bibfield  {journal}
  {\bibinfo  {journal} {New Journal of Physics}\ }\textbf {\bibinfo {volume}
  {21}},\ \bibinfo {pages} {063006} (\bibinfo {year} {2019})}\BibitemShut
  {NoStop}%
\bibitem [{\citenamefont {Dadhichi}\ \emph {et~al.}(2020)\citenamefont
  {Dadhichi}, \citenamefont {Kethapelli}, \citenamefont {Chajwa}, \citenamefont
  {Ramaswamy},\ and\ \citenamefont {Maitra}}]{dadhichi2020nonmutual}%
  \BibitemOpen
  \bibfield  {author} {\bibinfo {author} {\bibfnamefont {L.~P.}\ \bibnamefont
  {Dadhichi}}, \bibinfo {author} {\bibfnamefont {J.}~\bibnamefont
  {Kethapelli}}, \bibinfo {author} {\bibfnamefont {R.}~\bibnamefont {Chajwa}},
  \bibinfo {author} {\bibfnamefont {S.}~\bibnamefont {Ramaswamy}}, \ and\
  \bibinfo {author} {\bibfnamefont {A.}~\bibnamefont {Maitra}},\ }\href
  {\doibase 10.1103/PhysRevE.101.052601} {\bibfield  {journal} {\bibinfo
  {journal} {Phys. Rev. E}\ }\textbf {\bibinfo {volume} {101}},\ \bibinfo
  {pages} {052601} (\bibinfo {year} {2020})}\BibitemShut {NoStop}%
\bibitem [{\citenamefont {Fruchart}\ \emph {et~al.}(2020)\citenamefont
  {Fruchart}, \citenamefont {Hanai}, \citenamefont {Littlewood},\ and\
  \citenamefont {Vitelli}}]{fruchart2020phase}%
  \BibitemOpen
  \bibfield  {author} {\bibinfo {author} {\bibfnamefont {M.}~\bibnamefont
  {Fruchart}}, \bibinfo {author} {\bibfnamefont {R.}~\bibnamefont {Hanai}},
  \bibinfo {author} {\bibfnamefont {P.~B.}\ \bibnamefont {Littlewood}}, \ and\
  \bibinfo {author} {\bibfnamefont {V.}~\bibnamefont {Vitelli}},\ }\href
  {https://arxiv.org/abs/2003.13176} {\bibfield  {journal} {\bibinfo  {journal}
  {arXiv:2003.13176}\ } (\bibinfo {year} {2020})}\BibitemShut {NoStop}%
\bibitem [{\citenamefont {Saha}\ \emph {et~al.}(2020)\citenamefont {Saha},
  \citenamefont {Agudo-Canalejo},\ and\ \citenamefont
  {Golestanian}}]{saha2020scalar}%
  \BibitemOpen
  \bibfield  {author} {\bibinfo {author} {\bibfnamefont {S.}~\bibnamefont
  {Saha}}, \bibinfo {author} {\bibfnamefont {J.}~\bibnamefont
  {Agudo-Canalejo}}, \ and\ \bibinfo {author} {\bibfnamefont {R.}~\bibnamefont
  {Golestanian}},\ }\href {https://arxiv.org/abs/2005.07101v2} {\bibfield
  {journal} {\bibinfo  {journal} {arXiv:2005.07101}\ } (\bibinfo {year}
  {2020})}\BibitemShut {NoStop}%
\bibitem [{\citenamefont {You}\ \emph {et~al.}(2020)\citenamefont {You},
  \citenamefont {Baskaran},\ and\ \citenamefont
  {Marchetti}}]{you2020nonreciprocity}%
  \BibitemOpen
  \bibfield  {author} {\bibinfo {author} {\bibfnamefont {Z.}~\bibnamefont
  {You}}, \bibinfo {author} {\bibfnamefont {A.}~\bibnamefont {Baskaran}}, \
  and\ \bibinfo {author} {\bibfnamefont {M.~C.}\ \bibnamefont {Marchetti}},\
  }\href {https://arxiv.org/abs/2005.07684v2} {\bibfield  {journal} {\bibinfo
  {journal} {arXiv:2005.07684}\ } (\bibinfo {year} {2020})}\BibitemShut
  {NoStop}%
\bibitem [{\citenamefont {Kumar}\ \emph {et~al.}(2011)\citenamefont {Kumar},
  \citenamefont {Ramaswamy},\ and\ \citenamefont {Sood}}]{NKPRL}%
  \BibitemOpen
  \bibfield  {author} {\bibinfo {author} {\bibfnamefont {N.}~\bibnamefont
  {Kumar}}, \bibinfo {author} {\bibfnamefont {S.}~\bibnamefont {Ramaswamy}}, \
  and\ \bibinfo {author} {\bibfnamefont {A.~K.}\ \bibnamefont {Sood}},\ }\href
  {\doibase 10.1103/PhysRevLett.106.118001} {\bibfield  {journal} {\bibinfo
  {journal} {Phys. Rev. Lett.}\ }\textbf {\bibinfo {volume} {106}},\ \bibinfo
  {pages} {118001} (\bibinfo {year} {2011})}\BibitemShut {NoStop}%
\bibitem [{\citenamefont {Kumar}\ \emph {et~al.}(2015)\citenamefont {Kumar},
  \citenamefont {Soni}, \citenamefont {Ramaswamy},\ and\ \citenamefont
  {Sood}}]{NKPRE}%
  \BibitemOpen
  \bibfield  {author} {\bibinfo {author} {\bibfnamefont {N.}~\bibnamefont
  {Kumar}}, \bibinfo {author} {\bibfnamefont {H.}~\bibnamefont {Soni}},
  \bibinfo {author} {\bibfnamefont {S.}~\bibnamefont {Ramaswamy}}, \ and\
  \bibinfo {author} {\bibfnamefont {A.~K.}\ \bibnamefont {Sood}},\ }\href
  {\doibase 10.1103/PhysRevE.91.030102} {\bibfield  {journal} {\bibinfo
  {journal} {Phys. Rev. E}\ }\textbf {\bibinfo {volume} {91}},\ \bibinfo
  {pages} {030102} (\bibinfo {year} {2015})}\BibitemShut {NoStop}%
\bibitem [{\citenamefont {Kumar}\ \emph {et~al.}(2014)\citenamefont {Kumar},
  \citenamefont {Soni}, \citenamefont {Ramaswamy},\ and\ \citenamefont
  {Sood}}]{NatCom}%
  \BibitemOpen
  \bibfield  {author} {\bibinfo {author} {\bibfnamefont {N.}~\bibnamefont
  {Kumar}}, \bibinfo {author} {\bibfnamefont {H.}~\bibnamefont {Soni}},
  \bibinfo {author} {\bibfnamefont {S.}~\bibnamefont {Ramaswamy}}, \ and\
  \bibinfo {author} {\bibfnamefont {A.~K.}\ \bibnamefont {Sood}},\ }\href
  {http://dx.doi.org/10.1038/ncomms5688} {\bibfield  {journal} {\bibinfo
  {journal} {Nat Commun}\ }\textbf {\bibinfo {volume} {5}},\ \bibinfo {pages}
  {4688} (\bibinfo {year} {2014})}\BibitemShut {NoStop}%
\bibitem [{\citenamefont {Kumar}\ \emph {et~al.}(2019)\citenamefont {Kumar},
  \citenamefont {Gupta}, \citenamefont {Soni}, \citenamefont {Ramaswamy},\ and\
  \citenamefont {Sood}}]{trapping}%
  \BibitemOpen
  \bibfield  {author} {\bibinfo {author} {\bibfnamefont {N.}~\bibnamefont
  {Kumar}}, \bibinfo {author} {\bibfnamefont {R.~K.}\ \bibnamefont {Gupta}},
  \bibinfo {author} {\bibfnamefont {H.}~\bibnamefont {Soni}}, \bibinfo {author}
  {\bibfnamefont {S.}~\bibnamefont {Ramaswamy}}, \ and\ \bibinfo {author}
  {\bibfnamefont {A.~K.}\ \bibnamefont {Sood}},\ }\href {\doibase
  10.1103/PhysRevE.99.032605} {\bibfield  {journal} {\bibinfo  {journal} {Phys.
  Rev. E}\ }\textbf {\bibinfo {volume} {99}},\ \bibinfo {pages} {032605}
  (\bibinfo {year} {2019})}\BibitemShut {NoStop}%
\bibitem [{\citenamefont {Soni}\ \emph {et~al.}(2020)\citenamefont {Soni},
  \citenamefont {Kumar}, \citenamefont {Nambisan}, \citenamefont {Gupta},
  \citenamefont {Sood},\ and\ \citenamefont {Ramaswamy}}]{harshsoftmatter}%
  \BibitemOpen
  \bibfield  {author} {\bibinfo {author} {\bibfnamefont {H.}~\bibnamefont
  {Soni}}, \bibinfo {author} {\bibfnamefont {N.}~\bibnamefont {Kumar}},
  \bibinfo {author} {\bibfnamefont {J.}~\bibnamefont {Nambisan}}, \bibinfo
  {author} {\bibfnamefont {R.~K.}\ \bibnamefont {Gupta}}, \bibinfo {author}
  {\bibfnamefont {A.~K.}\ \bibnamefont {Sood}}, \ and\ \bibinfo {author}
  {\bibfnamefont {S.}~\bibnamefont {Ramaswamy}},\ }\href {\doibase
  10.1039/C9SM02552A} {\bibfield  {journal} {\bibinfo  {journal} {Soft Matter}\
  }\textbf {\bibinfo {volume} {16}},\ \bibinfo {pages} {7210} (\bibinfo {year}
  {2020})}\BibitemShut {NoStop}%
\bibitem [{nar()}]{narayan2010phase}%
  \BibitemOpen
  \href@noop {} {}\bibinfo {note} {V. Narayan, Phase Behaviour \& Dynamics of
  an Agitated Monolayer of Granular Rods, Ph.D. thesis, Indian Institute of
  Science(2010)}\BibitemShut {NoStop}%
\bibitem [{\citenamefont {Yamada}\ \emph {et~al.}(2003)\citenamefont {Yamada},
  \citenamefont {Hondou},\ and\ \citenamefont {Sano}}]{yamada2003coherent}%
  \BibitemOpen
  \bibfield  {author} {\bibinfo {author} {\bibfnamefont {D.}~\bibnamefont
  {Yamada}}, \bibinfo {author} {\bibfnamefont {T.}~\bibnamefont {Hondou}}, \
  and\ \bibinfo {author} {\bibfnamefont {M.}~\bibnamefont {Sano}},\ }\href
  {\doibase 10.1103/PhysRevE.67.040301} {\bibfield  {journal} {\bibinfo
  {journal} {Phys. Rev. E}\ }\textbf {\bibinfo {volume} {67}},\ \bibinfo
  {pages} {040301} (\bibinfo {year} {2003})}\BibitemShut {NoStop}%
\bibitem [{\citenamefont {Schneider}\ \emph {et~al.}(2012)\citenamefont
  {Schneider}, \citenamefont {Rasband},\ and\ \citenamefont
  {Eliceiri}}]{ImageJ}%
  \BibitemOpen
  \bibfield  {author} {\bibinfo {author} {\bibfnamefont {C.~A.}\ \bibnamefont
  {Schneider}}, \bibinfo {author} {\bibfnamefont {W.~S.}\ \bibnamefont
  {Rasband}}, \ and\ \bibinfo {author} {\bibfnamefont {K.~W.}\ \bibnamefont
  {Eliceiri}},\ }\href@noop {} {\bibfield  {journal} {\bibinfo  {journal}
  {Nature methods}\ }\textbf {\bibinfo {volume} {9}},\ \bibinfo {pages} {671}
  (\bibinfo {year} {2012})}\BibitemShut {NoStop}%
\bibitem [{\citenamefont {Stronge}(1994)}]{stronge}%
  \BibitemOpen
  \bibfield  {author} {\bibinfo {author} {\bibfnamefont {W.~J.}\ \bibnamefont
  {Stronge}},\ }\href {\doibase 10.1115/1.2901502} {\bibfield  {journal}
  {\bibinfo  {journal} {Journal of Applied Mechanics}\ }\textbf {\bibinfo
  {volume} {61}},\ \bibinfo {pages} {605} (\bibinfo {year} {1994})}\BibitemShut
  {NoStop}%
\bibitem [{\citenamefont {Humphrey}\ \emph {et~al.}(1996)\citenamefont
  {Humphrey}, \citenamefont {Dalke},\ and\ \citenamefont {Schulten}}]{VMD}%
  \BibitemOpen
  \bibfield  {author} {\bibinfo {author} {\bibfnamefont {W.}~\bibnamefont
  {Humphrey}}, \bibinfo {author} {\bibfnamefont {A.}~\bibnamefont {Dalke}}, \
  and\ \bibinfo {author} {\bibfnamefont {K.}~\bibnamefont {Schulten}},\
  }\href@noop {} {\bibfield  {journal} {\bibinfo  {journal} {Journal of
  Molecular Graphics}\ }\textbf {\bibinfo {volume} {14}},\ \bibinfo {pages}
  {33} (\bibinfo {year} {1996})}\BibitemShut {NoStop}%
\bibitem [{\citenamefont {Olafsen}\ and\ \citenamefont
  {Urbach}(2005)}]{olafsen2005two}%
  \BibitemOpen
  \bibfield  {author} {\bibinfo {author} {\bibfnamefont {J.~S.}\ \bibnamefont
  {Olafsen}}\ and\ \bibinfo {author} {\bibfnamefont {J.~S.}\ \bibnamefont
  {Urbach}},\ }\href {\doibase 10.1103/PhysRevLett.95.098002} {\bibfield
  {journal} {\bibinfo  {journal} {Phys. Rev. Lett.}\ }\textbf {\bibinfo
  {volume} {95}},\ \bibinfo {pages} {098002} (\bibinfo {year}
  {2005})}\BibitemShut {NoStop}%
\bibitem [{\citenamefont {Halperin}\ and\ \citenamefont
  {Nelson}(1978)}]{KTHNY2}%
  \BibitemOpen
  \bibfield  {author} {\bibinfo {author} {\bibfnamefont {B.~I.}\ \bibnamefont
  {Halperin}}\ and\ \bibinfo {author} {\bibfnamefont {D.~R.}\ \bibnamefont
  {Nelson}},\ }\href {\doibase 10.1103/PhysRevLett.41.519} {\bibfield
  {journal} {\bibinfo  {journal} {Phys. Rev. Lett.}\ }\textbf {\bibinfo
  {volume} {41}},\ \bibinfo {pages} {519} (\bibinfo {year} {1978})}\BibitemShut
  {NoStop}%
\bibitem [{\citenamefont {Brotto}\ \emph {et~al.}(2013)\citenamefont {Brotto},
  \citenamefont {Caussin}, \citenamefont {Lauga},\ and\ \citenamefont
  {Bartolo}}]{brotto2013}%
  \BibitemOpen
  \bibfield  {author} {\bibinfo {author} {\bibfnamefont {T.}~\bibnamefont
  {Brotto}}, \bibinfo {author} {\bibfnamefont {J.-B.}\ \bibnamefont {Caussin}},
  \bibinfo {author} {\bibfnamefont {E.}~\bibnamefont {Lauga}}, \ and\ \bibinfo
  {author} {\bibfnamefont {D.}~\bibnamefont {Bartolo}},\ }\href {\doibase
  10.1103/PhysRevLett.110.038101} {\bibfield  {journal} {\bibinfo  {journal}
  {Phys. Rev. Lett.}\ }\textbf {\bibinfo {volume} {110}},\ \bibinfo {pages}
  {038101} (\bibinfo {year} {2013})}\BibitemShut {NoStop}%
\bibitem [{sup()}]{supp}%
  \BibitemOpen
  \href@noop {} {}\bibinfo {note} {Supplementary videos are available at
  \url{https://www.dropbox.com/sh/od5lg00cxijk7b9/AAB3j6el7jFdaZ0q1xOo0Eqza?dl=0}}\BibitemShut
  {NoStop}%
\bibitem [{cur()}]{curlfoot}%
  \BibitemOpen
  \href@noop {} {}\bibinfo {note} {In the hydrodynamic treatment of
  \cite{NatCom} the rotation of the orientation field by the vorticity field is
  subleading in a gradient expansion and therefore neglected.}\BibitemShut
  {Stop}%
\bibitem [{\citenamefont {O'Byrne}\ and\ \citenamefont
  {Tailleur}(2020)}]{o2020lamellar}%
  \BibitemOpen
  \bibfield  {author} {\bibinfo {author} {\bibfnamefont {J.}~\bibnamefont
  {O'Byrne}}\ and\ \bibinfo {author} {\bibfnamefont {J.}~\bibnamefont
  {Tailleur}},\ }\href {\doibase 10.1103/PhysRevLett.125.208003} {\bibfield
  {journal} {\bibinfo  {journal} {Phys. Rev. Lett.}\ }\textbf {\bibinfo
  {volume} {125}},\ \bibinfo {pages} {208003} (\bibinfo {year}
  {2020})}\BibitemShut {NoStop}%
\bibitem [{\citenamefont {Narayan}\ \emph {et~al.}(2007)\citenamefont
  {Narayan}, \citenamefont {Ramaswamy},\ and\ \citenamefont
  {Menon}}]{VJScience}%
  \BibitemOpen
  \bibfield  {author} {\bibinfo {author} {\bibfnamefont {V.}~\bibnamefont
  {Narayan}}, \bibinfo {author} {\bibfnamefont {S.}~\bibnamefont {Ramaswamy}},
  \ and\ \bibinfo {author} {\bibfnamefont {N.}~\bibnamefont {Menon}},\ }\href
  {\doibase 10.1126/science.1140414} {\bibfield  {journal} {\bibinfo  {journal}
  {Science}\ }\textbf {\bibinfo {volume} {317}},\ \bibinfo {pages} {105}
  (\bibinfo {year} {2007})}\BibitemShut {NoStop}%
\bibitem [{\citenamefont {Eshelby}(1957)}]{eshelby1957determination}%
  \BibitemOpen
  \bibfield  {author} {\bibinfo {author} {\bibfnamefont {J.~D.}\ \bibnamefont
  {Eshelby}},\ }\href {https://doi.org/10.1098/rspa.1957.0133} {\bibfield
  {journal} {\bibinfo  {journal} {Proceedings of the Royal Society of London.
  Series A. Mathematical and Physical sciences}\ }\textbf {\bibinfo {volume}
  {241}},\ \bibinfo {pages} {376} (\bibinfo {year} {1957})}\BibitemShut
  {NoStop}%
\bibitem [{\citenamefont {Katira}\ \emph {et~al.}(2016)\citenamefont {Katira},
  \citenamefont {Mandadapu}, \citenamefont {Vaikuntanathan}, \citenamefont
  {Smit},\ and\ \citenamefont {Chandler}}]{katira}%
  \BibitemOpen
  \bibfield  {author} {\bibinfo {author} {\bibfnamefont {S.}~\bibnamefont
  {Katira}}, \bibinfo {author} {\bibfnamefont {K.~K.}\ \bibnamefont
  {Mandadapu}}, \bibinfo {author} {\bibfnamefont {S.}~\bibnamefont
  {Vaikuntanathan}}, \bibinfo {author} {\bibfnamefont {B.}~\bibnamefont
  {Smit}}, \ and\ \bibinfo {author} {\bibfnamefont {D.}~\bibnamefont
  {Chandler}},\ }\href {https://doi.org/10.7554/eLife.13150.002} {\bibfield
  {journal} {\bibinfo  {journal} {eLife}\ }\textbf {\bibinfo {volume} {5}},\
  \bibinfo {pages} {e13150} (\bibinfo {year} {2016})}\BibitemShut {NoStop}%
\bibitem [{\citenamefont {Landau}\ \emph {et~al.}(1986)\citenamefont {Landau},
  \citenamefont {Lifshitz}, \citenamefont {Kosevich},\ and\ \citenamefont
  {Pitaevskii}}]{landau}%
  \BibitemOpen
  \bibfield  {author} {\bibinfo {author} {\bibfnamefont {L.~D.}\ \bibnamefont
  {Landau}}, \bibinfo {author} {\bibfnamefont {E.~M.}\ \bibnamefont
  {Lifshitz}}, \bibinfo {author} {\bibfnamefont {A.~M.}\ \bibnamefont
  {Kosevich}}, \ and\ \bibinfo {author} {\bibfnamefont {L.~P.}\ \bibnamefont
  {Pitaevskii}},\ }\href {http://www.worldcat.org/isbn/075062633X} {\emph
  {\bibinfo {title} {Theory of Elasticity}}},\ \bibinfo {edition} {3rd}\ ed.\
  (\bibinfo  {publisher} {Butterworth-Heinemann},\ \bibinfo {year}
  {1986})\BibitemShut {NoStop}%
\bibitem [{\citenamefont {Ebeling}\ and\ \citenamefont
  {Schweitzer}(2002)}]{ebeling2002self}%
  \BibitemOpen
  \bibfield  {author} {\bibinfo {author} {\bibfnamefont {W.}~\bibnamefont
  {Ebeling}}\ and\ \bibinfo {author} {\bibfnamefont {F.}~\bibnamefont
  {Schweitzer}},\ }\href {https://arxiv.org/abs/cond-mat/0211606} {\bibfield
  {journal} {\bibinfo  {journal} {Nova Acta Leopoldina NF}\ }\textbf {\bibinfo
  {volume} {88}},\ \bibinfo {pages} {169} (\bibinfo {year} {2002})}\BibitemShut
  {NoStop}%
\bibitem [{\citenamefont {Romanczuk}\ \emph {et~al.}(2012)\citenamefont
  {Romanczuk}, \citenamefont {B{\"a}r}, \citenamefont {Ebeling}, \citenamefont
  {Lindner},\ and\ \citenamefont {Schimansky-Geier}}]{romanczuk2012active}%
  \BibitemOpen
  \bibfield  {author} {\bibinfo {author} {\bibfnamefont {P.}~\bibnamefont
  {Romanczuk}}, \bibinfo {author} {\bibfnamefont {M.}~\bibnamefont {B{\"a}r}},
  \bibinfo {author} {\bibfnamefont {W.}~\bibnamefont {Ebeling}}, \bibinfo
  {author} {\bibfnamefont {B.}~\bibnamefont {Lindner}}, \ and\ \bibinfo
  {author} {\bibfnamefont {L.}~\bibnamefont {Schimansky-Geier}},\ }\href
  {https://epjst.epj.org/articles/epjst/abs/2012/02/epjst202001/epjst202001.html}
  {\bibfield  {journal} {\bibinfo  {journal} {The European Physical Journal
  Special Topics}\ }\textbf {\bibinfo {volume} {202}},\ \bibinfo {pages}
  {1–162} (\bibinfo {year} {2012})}\BibitemShut {NoStop}%
\bibitem [{\citenamefont {Cates}\ and\ \citenamefont
  {Tailleur}(2013)}]{cates2013active}%
  \BibitemOpen
  \bibfield  {author} {\bibinfo {author} {\bibfnamefont {M.~E.}\ \bibnamefont
  {Cates}}\ and\ \bibinfo {author} {\bibfnamefont {J.}~\bibnamefont
  {Tailleur}},\ }\href {\doibase 10.1209/0295-5075/101/20010} {\bibfield
  {journal} {\bibinfo  {journal} {{EPL} (Europhysics Letters)}\ }\textbf
  {\bibinfo {volume} {101}},\ \bibinfo {pages} {20010} (\bibinfo {year}
  {2013})}\BibitemShut {NoStop}%
\bibitem [{\citenamefont {Saha}\ \emph {et~al.}(2014)\citenamefont {Saha},
  \citenamefont {Golestanian},\ and\ \citenamefont
  {Ramaswamy}}]{saha2014clusters}%
  \BibitemOpen
  \bibfield  {author} {\bibinfo {author} {\bibfnamefont {S.}~\bibnamefont
  {Saha}}, \bibinfo {author} {\bibfnamefont {R.}~\bibnamefont {Golestanian}}, \
  and\ \bibinfo {author} {\bibfnamefont {S.}~\bibnamefont {Ramaswamy}},\ }\href
  {https://doi.org/10.1103/PhysRevE.89.062316} {\bibfield  {journal} {\bibinfo
  {journal} {Physical Review E}\ }\textbf {\bibinfo {volume} {89}},\ \bibinfo
  {pages} {062316} (\bibinfo {year} {2014})}\BibitemShut {NoStop}%
\bibitem [{\citenamefont {Z{\"o}ttl}\ and\ \citenamefont
  {Stark}(2016)}]{zottl2016emergent}%
  \BibitemOpen
  \bibfield  {author} {\bibinfo {author} {\bibfnamefont {A.}~\bibnamefont
  {Z{\"o}ttl}}\ and\ \bibinfo {author} {\bibfnamefont {H.}~\bibnamefont
  {Stark}},\ }\href {https://doi.org/10.1088/0953-8984/28/25/253001} {\bibfield
   {journal} {\bibinfo  {journal} {Journal of Physics: Condensed Matter}\
  }\textbf {\bibinfo {volume} {28}},\ \bibinfo {pages} {253001} (\bibinfo
  {year} {2016})}\BibitemShut {NoStop}%
\bibitem [{\citenamefont {Liebchen}\ \emph {et~al.}(2017)\citenamefont
  {Liebchen}, \citenamefont {Marenduzzo},\ and\ \citenamefont
  {Cates}}]{liebchen2017phoretic}%
  \BibitemOpen
  \bibfield  {author} {\bibinfo {author} {\bibfnamefont {B.}~\bibnamefont
  {Liebchen}}, \bibinfo {author} {\bibfnamefont {D.}~\bibnamefont
  {Marenduzzo}}, \ and\ \bibinfo {author} {\bibfnamefont {M.~E.}\ \bibnamefont
  {Cates}},\ }\href {https://doi.org/10.1103/PhysRevLett.118.268001} {\bibfield
   {journal} {\bibinfo  {journal} {Physical Review Letters}\ }\textbf {\bibinfo
  {volume} {118}},\ \bibinfo {pages} {268001} (\bibinfo {year}
  {2017})}\BibitemShut {NoStop}%
\bibitem [{\citenamefont {Aranson}\ \emph {et~al.}(2007)\citenamefont
  {Aranson}, \citenamefont {Sokolov}, \citenamefont {Kessler},\ and\
  \citenamefont {Goldstein}}]{aranson2007model}%
  \BibitemOpen
  \bibfield  {author} {\bibinfo {author} {\bibfnamefont {I.~S.}\ \bibnamefont
  {Aranson}}, \bibinfo {author} {\bibfnamefont {A.}~\bibnamefont {Sokolov}},
  \bibinfo {author} {\bibfnamefont {J.~O.}\ \bibnamefont {Kessler}}, \ and\
  \bibinfo {author} {\bibfnamefont {R.~E.}\ \bibnamefont {Goldstein}},\ }\href
  {https://journals.aps.org/pre/abstract/10.1103/PhysRevE.75.040901} {\bibfield
   {journal} {\bibinfo  {journal} {Physical Review E}\ }\textbf {\bibinfo
  {volume} {75}},\ \bibinfo {pages} {040901} (\bibinfo {year}
  {2007})}\BibitemShut {NoStop}%
\bibitem [{\citenamefont {Maitra}\ \emph {et~al.}(2020)\citenamefont {Maitra},
  \citenamefont {Srivastava}, \citenamefont {Marchetti}, \citenamefont
  {Ramaswamy},\ and\ \citenamefont {Lenz}}]{maitra2020swimmer}%
  \BibitemOpen
  \bibfield  {author} {\bibinfo {author} {\bibfnamefont {A.}~\bibnamefont
  {Maitra}}, \bibinfo {author} {\bibfnamefont {P.}~\bibnamefont {Srivastava}},
  \bibinfo {author} {\bibfnamefont {M.~C.}\ \bibnamefont {Marchetti}}, \bibinfo
  {author} {\bibfnamefont {S.}~\bibnamefont {Ramaswamy}}, \ and\ \bibinfo
  {author} {\bibfnamefont {M.}~\bibnamefont {Lenz}},\ }\href
  {https://doi.org/10.1103/PhysRevLett.124.028002} {\bibfield  {journal}
  {\bibinfo  {journal} {Physical Review Letters}\ }\textbf {\bibinfo {volume}
  {124}},\ \bibinfo {pages} {028002} (\bibinfo {year} {2020})}\BibitemShut
  {NoStop}%
\bibitem [{\citenamefont {Oseen}(1910)}]{oseen1910uber}%
  \BibitemOpen
  \bibfield  {author} {\bibinfo {author} {\bibfnamefont {C.~W.}\ \bibnamefont
  {Oseen}},\ }\href
  {https://www.pdf-archive.com/2020/02/23/untitled-pdf-document-3} {\bibfield
  {journal} {\bibinfo  {journal} {Arkiv Mat., Astron. och Fysik}\ }\textbf
  {\bibinfo {volume} {6}},\ \bibinfo {pages} {1} (\bibinfo {year}
  {1910})}\BibitemShut {NoStop}%
\bibitem [{\citenamefont {Stokes}(1850)}]{stokes1850effect}%
  \BibitemOpen
  \bibfield  {author} {\bibinfo {author} {\bibfnamefont {G.~G.}\ \bibnamefont
  {Stokes}},\ }\href
  {https://babel.hathitrust.org/cgi/pt?id=mdp.39015012112531&view=1up&seq=208}
  {\bibfield  {journal} {\bibinfo  {journal} {Trans. Camb. Phil. Soc}\ }\textbf
  {\bibinfo {volume} {9}},\ \bibinfo {pages} {8} (\bibinfo {year}
  {1850})}\BibitemShut {NoStop}%
\bibitem [{\citenamefont {Gradshte{\u\i}n}\ and\ \citenamefont
  {Ryzhik}(2014)}]{gradshteyn2014table}%
  \BibitemOpen
  \bibfield  {author} {\bibinfo {author} {\bibfnamefont {I.~S.}\ \bibnamefont
  {Gradshte{\u\i}n}}\ and\ \bibinfo {author} {\bibfnamefont {I.~M.}\
  \bibnamefont {Ryzhik}},\ }\href
  {https://www.elsevier.com/books/table-of-integrals-series-and-products/jeffrey/978-0-08-047111-2}
  {\emph {\bibinfo {title} {Table of integrals, series, and products}}}\
  (\bibinfo  {publisher} {Academic press},\ \bibinfo {year} {2014})\BibitemShut
  {NoStop}%
\bibitem [{len()}]{lengthscalefoot}%
  \BibitemOpen
  \href@noop {} {}\bibinfo {note} {At large wavevectors ${\bf q}$ where the
  $v_0$ term can be ignored, the Fourier transform ${\bf U}_{\bf q} \sim f/\mu
  q^2$, i.e., $f/\mu$ as a source for the displacement field enters in a manner
  reminiscent of the Burgers vector of a dislocation.}\BibitemShut {Stop}%
\bibitem [{\citenamefont {Reynolds}(1885)}]{reynolds1885lvii}%
  \BibitemOpen
  \bibfield  {author} {\bibinfo {author} {\bibfnamefont {O.}~\bibnamefont
  {Reynolds}},\ }\href@noop {} {\bibfield  {journal} {\bibinfo  {journal} {The
  London, Edinburgh, and Dublin Philosophical Magazine and Journal of Science}\
  }\textbf {\bibinfo {volume} {20}},\ \bibinfo {pages} {469} (\bibinfo {year}
  {1885})}\BibitemShut {NoStop}%
\bibitem [{\citenamefont {Yoshinaga}\ \emph {et~al.}(2010)\citenamefont
  {Yoshinaga}, \citenamefont {Joanny}, \citenamefont {Prost},\ and\
  \citenamefont {Marcq}}]{yoshinaga2010polarity}%
  \BibitemOpen
  \bibfield  {author} {\bibinfo {author} {\bibfnamefont {N.}~\bibnamefont
  {Yoshinaga}}, \bibinfo {author} {\bibfnamefont {J.-F.}\ \bibnamefont
  {Joanny}}, \bibinfo {author} {\bibfnamefont {J.}~\bibnamefont {Prost}}, \
  and\ \bibinfo {author} {\bibfnamefont {P.}~\bibnamefont {Marcq}},\ }\href
  {https://doi.org/10.1103/PhysRevLett.105.238103} {\bibfield  {journal}
  {\bibinfo  {journal} {Physical Review Letters}\ }\textbf {\bibinfo {volume}
  {105}},\ \bibinfo {pages} {238103} (\bibinfo {year} {2010})}\BibitemShut
  {NoStop}%
\bibitem [{\citenamefont {Peleg}\ \emph {et~al.}(2018)\citenamefont {Peleg},
  \citenamefont {Peters}, \citenamefont {Salcedo},\ and\ \citenamefont
  {Mahadevan}}]{peleg2018collective}%
  \BibitemOpen
  \bibfield  {author} {\bibinfo {author} {\bibfnamefont {O.}~\bibnamefont
  {Peleg}}, \bibinfo {author} {\bibfnamefont {J.~M.}\ \bibnamefont {Peters}},
  \bibinfo {author} {\bibfnamefont {M.~K.}\ \bibnamefont {Salcedo}}, \ and\
  \bibinfo {author} {\bibfnamefont {L.}~\bibnamefont {Mahadevan}},\ }\href
  {https://doi.org/10.1038/s41567-018-0262-1} {\bibfield  {journal} {\bibinfo
  {journal} {Nature Physics}\ }\textbf {\bibinfo {volume} {14}},\ \bibinfo
  {pages} {1193} (\bibinfo {year} {2018})}\BibitemShut {NoStop}%
\bibitem [{\citenamefont {Theveneau}\ \emph {et~al.}(2013)\citenamefont
  {Theveneau}, \citenamefont {Steventon}, \citenamefont {Scarpa}, \citenamefont
  {Garcia}, \citenamefont {Trepat}, \citenamefont {Streit},\ and\ \citenamefont
  {Mayor}}]{theveneau2013chase}%
  \BibitemOpen
  \bibfield  {author} {\bibinfo {author} {\bibfnamefont {E.}~\bibnamefont
  {Theveneau}}, \bibinfo {author} {\bibfnamefont {B.}~\bibnamefont
  {Steventon}}, \bibinfo {author} {\bibfnamefont {E.}~\bibnamefont {Scarpa}},
  \bibinfo {author} {\bibfnamefont {S.}~\bibnamefont {Garcia}}, \bibinfo
  {author} {\bibfnamefont {X.}~\bibnamefont {Trepat}}, \bibinfo {author}
  {\bibfnamefont {A.}~\bibnamefont {Streit}}, \ and\ \bibinfo {author}
  {\bibfnamefont {R.}~\bibnamefont {Mayor}},\ }\href
  {https://www.nature.com/articles/ncb2772} {\bibfield  {journal} {\bibinfo
  {journal} {Nature cell biology}\ }\textbf {\bibinfo {volume} {15}},\ \bibinfo
  {pages} {763} (\bibinfo {year} {2013})}\BibitemShut {NoStop}%
\bibitem [{\citenamefont {Mayett}\ \emph {et~al.}(2017)\citenamefont {Mayett},
  \citenamefont {Bitten}, \citenamefont {Das},\ and\ \citenamefont
  {Schwarz}}]{mayett2017chase}%
  \BibitemOpen
  \bibfield  {author} {\bibinfo {author} {\bibfnamefont {D.}~\bibnamefont
  {Mayett}}, \bibinfo {author} {\bibfnamefont {N.}~\bibnamefont {Bitten}},
  \bibinfo {author} {\bibfnamefont {M.}~\bibnamefont {Das}}, \ and\ \bibinfo
  {author} {\bibfnamefont {J.~M.}\ \bibnamefont {Schwarz}},\ }\href
  {https://journals.aps.org/pre/abstract/10.1103/PhysRevE.96.032407} {\bibfield
   {journal} {\bibinfo  {journal} {Physical Review E}\ }\textbf {\bibinfo
  {volume} {96}},\ \bibinfo {pages} {032407} (\bibinfo {year}
  {2017})}\BibitemShut {NoStop}%
\end{thebibliography}%

\end{document}